\documentclass[prd,twocolumn,superscriptaddress,floatfix,amsmath,amssymb,amsfonts,nofootinbib,longbibliography]{revtex4-2}
\usepackage{float} 
\usepackage{scalerel}
\usepackage[normalem]{ulem}
\usepackage[english]{babel}
\usepackage{graphicx}
\usepackage{dcolumn}
\usepackage{bm}
\usepackage{blindtext}
\usepackage{verbatim}
\usepackage{relsize}
\usepackage{mathrsfs}
\usepackage{musicography}
\usepackage{amsmath}
\usepackage{blindtext}
\usepackage{cancel}
\usepackage{physics}
\usepackage{epstopdf}
\usepackage{mathtools}
\usepackage{blindtext}
\usepackage{tensor}
\usepackage{color}
\usepackage[usenames,dvipsnames]{pstricks}
\usepackage{epsfig}
\usepackage{pst-grad} 
\usepackage{pst-plot} 
\usepackage{hyperref}
\usepackage{verbatim}
\usepackage{slashed}
\usepackage{dsfont}
\usepackage{amsmath,amssymb,amsthm}  
\usepackage{bbold}
\usepackage{lipsum}

\newcommand{\ii}{\mathrm{i}}

\newcommand{\s}{\hat{\sigma}}

\newcommand{\lb}{\left(}
\newcommand{\rb}{\right)}

\newcommand{\bldk}{\mathbf{k}}

\allowdisplaybreaks[1] 

\usepackage{cleveref}

\begin{document}

\title{Trajectory-Protected Quantum Computing} 

\author{Barbara \v{S}oda}
\email{bsoda@perimeterinstitute.ca}

\affiliation{Department of Physics, Faculty of Science, University of Zagreb, 10000 Zagreb, Croatia}

\affiliation{Perimeter Institute for Theoretical Physics, Waterloo, Ontario, N2L 2Y5, Canada}

\author{Pierre-Antoine Graham}

\affiliation{Perimeter Institute for Theoretical Physics, Waterloo, Ontario, N2L 2Y5, Canada}
\affiliation{Department of Physics and Astronomy, University of Waterloo, Waterloo, Ontario, N2L 3G1, Canada}

\author{T. Rick Perche}
\email{rick.perche@su.se}
\affiliation{Nordita,
KTH Royal Institute of Technology and Stockholm University,
Hannes Alfvéns väg 12, 23, SE-106 91 Stockholm, Sweden}

\author{Gurpahul Singh}

\affiliation{Perimeter Institute for Theoretical Physics, Waterloo, Ontario, N2L 2Y5, Canada}
\affiliation{Department of Physics and Astronomy, University of Waterloo, Waterloo, Ontario, N2L 3G1, Canada}

\begin{abstract}
We introduce a novel method that simultaneously isolates a quantum computer from decoherence and enables the controlled implementation of computational gates. We demonstrate a quantum computing model that utilizes a qubit’s motion to protect it from decoherence. We model a qubit interacting with a quantum field via the standard light-matter interaction model: an Unruh-DeWitt detector, i.e. the qubit, follows a prescribed classical trajectory while interacting with a scalar quantum field.  We switch off the rotating-wave terms, i.e. the resonant transitions using the technique of acceleration-induced transparency which eliminates the dominant decoherence channels by controlling the qubit's trajectory. We are able to perform one qubit gates by stimulating the counter-rotating wave terms (i.e. the non-resonant transitions) and two qubit gates by extracting the entanglement from the quantum field prepared in a squeezed state. Finally, we discuss the fundamental limits on quantum error protection: on the trade-off between isolating a quantum computer from decoherence, and the speed with which entangling gates may be applied, comparable to the Eastin-Knill theorem for quantum error correction.
\end{abstract}

\maketitle

\textbf{Introduction.} In this work, we propose a quantum computing model which aims to reconcile seemingly contradictory requirements for a useful quantum computer: a) the need to isolate it from its environment, in order to protect it from decoherence,
and b) the necessity of interacting with it, in order to perform computational gates. Our framework is fundamentally distinct from conventional quantum error correction models.

The contradictory requirements suggest we must gain control over a quantum computer's interactions. Here, without loss of generality, we will focus on qubits and their interactions with each other and their environment. A qubit may be realized as a piece of matter, such as an atom \cite{RydbergatomsQC} or an ion \cite{Trappedions}, or a photon \cite{PhotonicQC}. By performing a sufficient number of experiments to characterize the qubit and its environment, one may determine a qubit's free Hamiltonian $H_0$ and its interaction Hamiltonian $H_{int}$ are to a fine enough level of detail.

Typically, we find that a qubit interacts more strongly via some channels of the interaction Hamiltonian, and less so through others \cite{OpenQS1, JaynesCummings}. This means separating the interactions into strong and weak, or in a familiar terminology, into resonant and non-resonant interactions. In the models of light-matter interactions, these correspond to rotating wave terms, and counter-rotating wave terms. Schematically:
\begin{equation}
H_{int}=H_{int}^{(strong)}+H_{int}^{(weak)}.
\end{equation}

To control decoherence and computational channels of a qubit, we propose the following: suppressing the strong interactions, ideally even to exact zero, via an external classical control over parameters of the interaction Hamiltonian. This  gives us the ability to suppress decoherence. At the same time, we propose modulating the strength of weak interactions by controlling the quantum state of the quantum field, to control the application of quantum gates. Then, the typical sources of decoherence, namely the strong, or resonant, transitions are eliminated, while the weak ones contribute negligibly to decoherence, and can be amplified at will by controlling the state of the field, in order to perform computational gates.

As a particular representation of the general procedure, we examine in detail the interaction between a qubit and a quantum field, employing the Unruh–DeWitt detector model \cite{Unruh1976}, a standard framework for describing light–matter interactions. In our framework, the qubit is modeled as an Unruh–DeWitt detector ~\cite{Achim_Eduardo_2013,aspling2024}. It is a two-level system following a prescribed classical trajectory and interacting with a scalar quantum field confined in a cavity, which serves as its environment. This interaction provides a good approximation to, for example, an atomic orbital or a charged particle coupled to the electromagnetic field. Within this model, the classical trajectory acts as an externally controlled parameter of the Hamiltonian.

Our goal is to clearly separate the interaction channels that typically cause strong decoherence, namely, the resonant terms, from those we propose to use for gate implementation, i.e., the remaining non-resonant terms. While the latter are generally weak, they can be externally stimulated. As a first step, we analyze the interactions between the qubits and their environment. Once we complete that step, we will be prepared to describe how computational gates are implemented.

\begin{figure}[h!]
    \centering
    \includegraphics[width=0.45\textwidth]{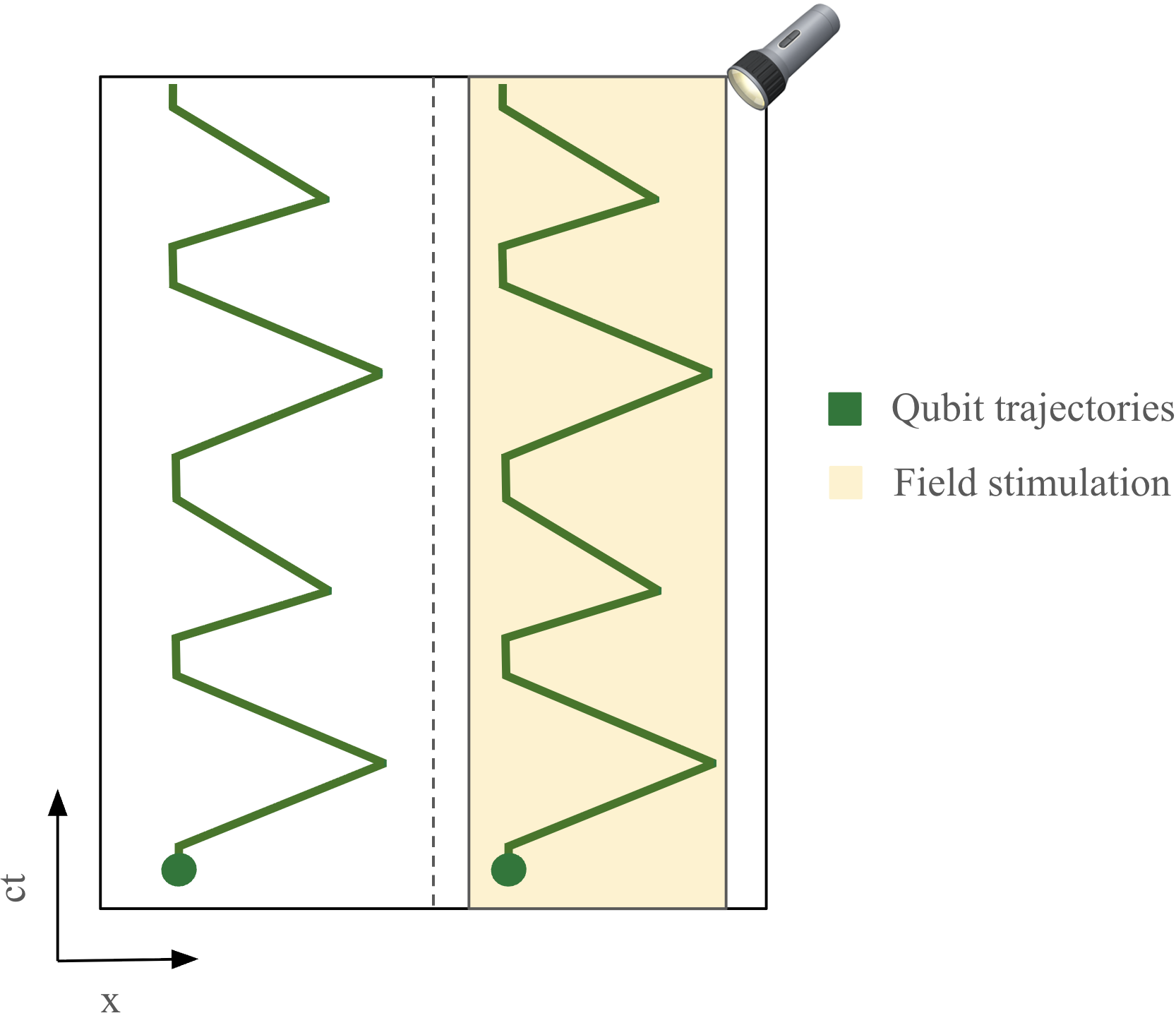}
    \caption{A schematic depiction of our quantum computing setup: qubits are placed in trajectories, and they are following prescribed, externally controlled trajectories. We can also control the state of the quantum field in its cavity environment by producing a coherent state,  represented here schematically by shining a laser onto the right qubit's cavity, or a squeezed state.}
    \label{fig:qubit_trajectory_cavity}
\end{figure}

\textbf{Light-matter interactions.} A system consisting of an atom interacting with the electromagnetic field in a cavity is well approximated \cite{Pozas-Kerstjens:2015,richard} by the interaction of a qubit with harmonic oscillators, each corresponding to a cavity mode. In this context, a qubit's two states represent the atom's relevant energy levels, ground and excited. We denote the creation and annihilation operators associated with the cavity modes by $\hat{a}_{n}^\dagger$, $\hat{a}_{n}$, and the discrete index $n$ labels the modes. The Hamiltonian of the system in the Schr\"odinger picture can be written as the sum of free Hamiltonians of the field and the detector, and the interaction Hamiltonian, $\hat{H}=\hat{H}_0^{\textup{(field)}}+\hat{H}_0^{\textup{(det.)}}+\hat{H}_{int}$:
\begin{align}\label{2QH}
    \hat{H} = &\ \frac{\text{d} t(\tau)}{\text{d} \tau}\sum_{n} |\bm k_n| \hat{a}^\dagger_n \hat{a}_n + \Omega  \sigma^+ \s^- \nonumber \\
    &+ \sum_{n} \lambda_n (\s^+ + \s^-)(\hat{a}_n e^{i\bm{k}_n \cdot \bm{x}_j(\tau)} + \hat{a}^\dagger_n e^{-i\bm{k}_n \cdot \bm{x}_j(\tau)}),
\end{align}
where $\Omega$ is the energy gap and $|\bm k_n|$ are the energies of each mode of the electromagnetic field. $\hat{\sigma}^\pm$ are the lowering and raising operators in the qubit system, and $\lambda_n$ are coupling constants that control the intensity of the interaction of the qubit with each cavity mode. We add the  factor of $\text{d} t(\tau)/\text{d}\tau$ to account for the possibility of a significant Doppler effect in the case of relativistic motion of a qubit creating a significant difference between proper time $\tau$ and coordinate time $t$. In the interaction picture, the interaction Hamiltonian takes the shape:

\begin{align}\label{eq:HI_relat}
\hat{H}_{int}(\tau) 
   & = \sum_n \left[
        \lambda_n e^{\ii \Omega \tau+\ii |{\bm k_n}| t(\tau) - \ii \bm{k}_n \cdot \bm{x}(\tau)}\hat{\sigma}^+\hat{a}^\dagger_{n} \right. + \notag \\
    &\quad \left.
        + \lambda_n e^{- \ii \Omega \tau+\ii |{\bm k_n}| t(\tau) - \ii \bm{k}_n \cdot \bm{x}(\tau)}
        \hat{\sigma}^- 
        + \textup{h.c.} \right].
\end{align}

The parameters $\bm{x}(\tau)$ and $t(\tau)$ denote the spacetime trajectory of the center of mass of the atom in the laboratory frame, parametrized by its proper time $\tau$. This Hamiltonian incorporates the classical trajectory of the qubit as a classical parameter of a drive term. 

The dynamics contains two qualitatively distinct contributions. The operators $\hat{\sigma}^-\hat{a}_{n}^{\dagger}$ and $\hat{\sigma}^+\hat{a}_{n}$ represent resonant terms, which create a mode excitation of the field while lowering the electron's energy level in the atom, or the other way around. In literature, these are typically also referred to as the rotating-wave terms. The remaining terms are non-resonant, $\hat{\sigma}^-\hat{a}_{n}$ and $\hat{\sigma}^+\hat{a}_{n}^{\dagger}$ and typically have a significantly smaller effect in the system's dynamics. For this reason, the rotating wave approximation (RWA) is usually employed to describe the system's dynamics, which neglects the non-resonant terms (often referred to as the non-rotating wave terms).


\textbf{Shielding from decoherence: acceleration-induced transparency.} Acceleration-induced transparency \cite{barbara} suppresses certain resonant transitions between a two-level system and a quantum field to exactly zero at first order in the coupling constant. By controlling the qubits’ trajectories, we harness this effect to suppress the dominant decoherence channels.

In more details, the transparency is achieved by utilizing interaction Hamiltonian's dependence on the trajectory, at first order expansion of the total interaction picture unitary operator ~\cite{barbara}:
\begin{align}\label{eq:Unitary_Op}
     \hat{U}(0, T)&= \mathcal{T} \exp(-i \int_0^{ T} \hat{H}_I(\tau') \text{d}\tau'),\\
     &\approx 1 - i \int_0^{ T} \hat{H}_I(\tau') \text{d}\tau'. 
\end{align}

The resonant and non-resonant terms can be neatly delineated in their different time dependence, as shown in the Hamiltonian \cref{eq:HI_relat}, which directly leads to differences in the time evolution at first order in \cref{eq:Unitary_Op}. The key insight is to identify  that the probability for transitions via the resonant terms are proportional to the square of one type of time integral, which we call $\abs{I^-}^2$ due to the minus sign between phases accumulated by the detector and the field, while the non-resonant terms are proportional to another time integral $\abs{I^+}^2$, for analogous reasons. They are defined as:
\begin{equation}\label{eq:Ipm}
    I^{\pm}\lb\Omega,|\bm k_n|\rb = \int \dd \tau\, e^{\ii \Omega\tau \pm \ii \lb |\bm k_n|t(\tau) - \bm k_n \cdot \bm x(\tau) \rb}.
\end{equation}

 In the language of signal processing, the changes in motion are causing a chirp, and the time integrals $I^{\pm}\lb\Omega,|\bm k_n|\rb$ are sensitive to the frequency content of that chirp. Exact transparency is achieved when, for the same trajectory, we have $\abs{I^-}=0$ and $\abs{I^+}\neq0$. Then, for the qubit energy gap $\Omega$ and the field mode $\bm k_n$, the resonant transitions are off, while the non-resonant ones are still present, though they may be small in magnitude, depending on the details of the qubit's trajectory.

While this first-order analysis provides valuable intuition, we will later consider a higher order expansion of the unitary operator, which is necessary to understand the full lowest order single- and two-qubit gates.

We aim to use the effect of transparency to suppress the resonant contributions, proportional to $I^-$, to reduce the dominant source of decoherence affecting qubits. For instance, it is known that these are the main source of decoherence for atomic systems undergoing inertial motion ~\cite{Pozas2016, MartinMartinez2013}. The rate of decoherence depends on both resonant and off-resonant contributions, which can be seen in the probability of the qubit changing state from ground to excited via interaction with a scalar field in ~\cite{barbara}:
\begin{equation}
	\mathcal{P}(\bldk) \label{eq:decoherence_probability}
\propto \left( \langle\hat{a}_\bldk^\dagger \hat{a}_\bldk \rangle 
		\abs{I^+ + I^-}^2 + \abs{I^+}^2 \right).
\end{equation}

Their relative magnitude matters in this analysis, and we refer to e.g. paper ~\cite{Ahmadzadegan2018} as indication of how difficult it is to enlarge the off-resonant contribution, typically requiring non-smoothness of the qubit's trajectory.

In ~\cite{barbara}, the existence of the acceleration-induced transparency was proven for a certain class of trajectories. However, the transparency  was only achieved for trajectories that extend both to spatial and time-like infinity. In this work, we demonstrate examples of trajectories that lead to transparencies and are compact in both space and time. This is a crucial improvement, since the qubits' trajectories are expected to be realized in a  laboratory setting, for a realistic model of a quantum computer.

Additionally, in this work we situate the qubits in cavities, as depicted in Fig.~\ref{fig:qubit_trajectory_cavity}. This is in order to reduce the complexity of protection against resonant interaction with the entire continuum of field modes to only a discrete number of them. Which field modes are present in the cavity ultimately depends on its shape. As long as the cavity's modes correspond to the transparent modes for the qubit, resonant transitions are effectively switched off. In earlier studies, we found that it is likely possible to find trajectories with multiple transparencies, i.e. trajectories that effectively shut down resonant transitions via more than one mode of the quantum field. However, in the version presented here, we focus on only one mode of the field. Physically, this would correspond to the assumption that there is one dominant mode of decoherence we are trying to suppress, and the effect of the remaining modes is negligible. The more general strategy would be to find trajectories transparent to some modes, and then leave the discussion of cavity shape-engineering to fit the transparency modes, see e.g. \cite{Karpov2023ModeOnDemand,Pal2023InverseMicroresonators}. We assume, even in the one-mode approximation, a high Q-factor for the cavity, such that energy loss is small, and the resonant peaks are narrow.

Other strategies to mitigate decoherence from electromagnetic coupling include dynamical decoupling~\cite{Viola1999} and decoherence-free subspaces~\cite{Lidar1998}. In current superconducting-qubit architectures, coherence times are on the order of \(T_2 \sim 1\)~ms~\cite{Place2021}, which—together with typical single-qubit gate durations of \(\sim 100\)~ns—corresponds to a budget of \(\mathcal{O}(10^4)\) gate operations per coherence time. In order to improve that, we suggest a new strategy.

We first outline the constraints required of admissible trajectories, and the construction of such trajectories. Then we will explain the application of single- and two-qubit gates, and finally discuss the advantages and limitations of this model, together with a new view of the Eastin-Knill theorem.

\textbf{Externally controlled qubit trajectories.} Our goal is to find trajectories transparent for the relevant mode of the cavity, making the qubits transparent to the resonant terms of the electromagnetic interaction. 
Admissible trajectories must satisfy three constraints: (a) cyclicity---each qubit returns to its initial position at the end of every acceleration–deceleration cycle, (b) concatenability—cycles can be repeated while preserving transparency, and (c) controllable off-resonant coupling—the remaining non-resonant terms are nonzero but small, enabling gate implementation by stimulating or appropriately preparing the field state.

In the continuation of this paper, and Supplementary Information, we demonstrate such trajectories.

\textbf{Periodicity constraint.} The computation scheme proposed here divides time in computation steps of duration $T$. The trajectory design we considered ensures that the qubit’s position, $\bm{x}(\tau)$, is periodic after each cycle, i.e., its position at corresponding times in successive cycles remains the same. It is also necessary that the Hamiltonian is periodic within each cycle. Enforcing such a periodicity constraint guarantees that the phase at the beginning and end of each cycle match, preventing any spurious phase accumulation in the calculation of both one-qubit and two-qubit gates.

\begin{figure}
    \centering
\includegraphics[width=0.7\linewidth]{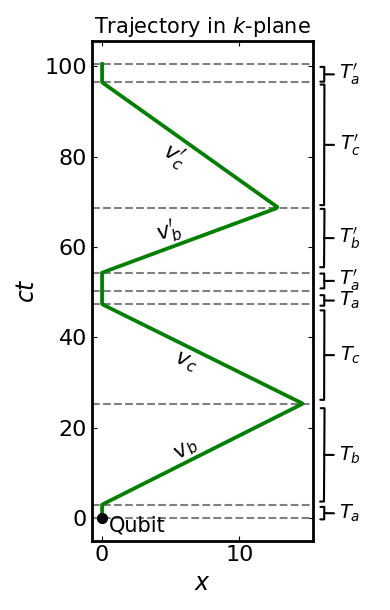}
    \caption{Plot of a trajectory that satisfies both transparency and periodicity constraints. It consists of a concatenation of inertial trajectories with different velocities, e.g. $v_c = 0.66,v'_c=0.46$, and the rest are explained in Supplementary Information.}
    \label{fig:trajectory_vamp}
\end{figure}

An example of such a trajectory, and the details of its transparency are given in Fig.~\ref{fig:trajectory_vamp} and Supplementary Information.

To perform quantum computations, we then consider multiple qubits undergoing identical transparent trajectories in cycles. While they are shielded from decoherence, we apply the quantum computational gates. However, while transparency shields the qubits from decoherence, it makes them transparent to all resonant interactions, which are commonly used to apply quantum gates and control the qubits. This requires us to also think of implementations of quantum computing from a new perspective, where the (weak) non-resonant interactions are used to implement local unitaries and entangling gates, which we discuss in the single-qubit gate section. 

\textbf{Single-qubit gates.} Having suppressed the resonant terms by using acceleration-induced transparency, we use the non-resonant terms to apply gates to a qubit. These terms cause negligible decoherence and are frequently neglected, as they are significantly smaller than the resonant terms. However, we can stimulate them by the presence of excitations of the quantum field (e.g. photons for the electromagnetic field). Practically, in a laboratory, that means illuminating the qubit with a laser field to stimulate non-resonant transitions. Therefore, colloquially, the switching of a laser, directly corresponds to the switching of computational single-qubit gates.

To demonstrate the single-qubit gates, we start from the expressions for the interaction Hamiltonian $\hat{H}_{int}$ in the interaction picture \cref{2QH}, and obtain an effective Hamiltonian~\cite{bukov_universal_2015}. Details of the derivation of the effective single-qubit Hamiltonian are given in the Supplementary Information . Importantly, for single-qubit gates, we note that the naive expansion to only the first order in the coupling constant is not enough, as there is a second order term that acts on each qubit individually. This higher order term becomes relevant whenever the field is in a highly excited state e.g. a coherent state of the field with $\expval{\hat{a}^{\dagger}\hat{a}} \gg 1$ can lead to a great amplification of terms of the form $\lambda^2 \hat{a}^\dagger\hat{a}$. 

In more details, the effective Hamiltonian is given by  $\hat{H}_{\rm eff} = \hat{H}_1 + \hat{H}_2$, corresponding to $\hat{U}(0, T)$ consisting of the following terms: 
\begin{align}\label{eq:H1}
     &\hat{H}_1 =  \frac{\lambda}{T} \sum_j \left(\s_j^+ \hat{a}^\dagger I_{j}^+ + \s_j^+ \hat{a} I_{j}^- + \text{h.c.}\right) +\ii\frac{\lambda^2}{T}\sum_j \hat{\delta}_j  \s_j^z,\\\label{eq:H2}
     &\hat{H}_2 = \ii\frac{\lambda^2}{T}\sum_{j> k} \sum_{s_i \in \{+,-\}} (A_{s_1 s_2}^{(jk)}+A_{s_1 s_2}^{(kj)}) \s_j^{s_1}\s_k^{s_2}, 
 \end{align}
 
where $j$ ($k$) denotes a $j$-th ($k$-th) qubit, $T$ is the time duration of once cycle of the qubit, following our previous discussion on periodic trajectories in space and time, and $I^{\pm}$ are defined in eq.~\eqref{eq:Ipm} as time integrals of phase functions which arise from the qubits' trajectories. $\hat{\delta}_{j}$ are operators which contain  products of two field operators, and are therefore non-negligible, since their expected value may stimulate the second order in coupling constant. The numbers $A_{s_1,s_2}^{(jk)}$ depend on the details of the time integrals, given in detail in the Supplementary Information.
 
While $\hat{H}_1$ is used to implement single detector gates for high stimulation with a cavity coherent state, $\hat{H}_2$ provides cavity independent interaction between the detectors through the matrix $A_{s_1 s_2}$. Interaction between the cavity and detectors induces a detuning of $\Omega$ parameterized by $\hat{\delta}_j$ which depends on the state of the cavity. This detuning can only become comparable to an order 1 effect for very specific trajectories where $A_{s_1 s_2}$ become large enough to compensate for $\lambda^2 \ll 1$. This is unlikely to happen, and our goal here is in fact the opposite: leave the detuning as small as possible.

Now, we will assume that we are using a trajectory of the qubit which sets the resonant contribution to zero, via the setting of time integral $I^-\left(\Omega, \mathbf{k}\right)$ to zero, given a fixed energy gap $\Omega$ for the qubits, and a chosen mode of the field $\mathbf{k}$. Additionally, to stimulate the non-resonant terms, we will assume that the field is in a coherent state $\ket{\alpha}$, for the transparent mode $\mathbf{k}$. And in order to stimulate the small coupling constant and the generically small contribution from the $I^+$ integral, we typically need to have $\abs{\alpha}\gg 1$. Then, we are then left with the effective single-qubit Hamiltonian:

 \begin{align*}\label{eq:Transparent_H1}
     \hat{H}_1 
     &= \frac{\lambda}{T} \sum_j  \left(\mathrm{Re}[\alpha^\star I_{j}^+]\hat{\sigma}^x_j - \mathrm{Im}[\alpha^\star I_{j}^+] \hat{\sigma}^y_j -i \lambda \delta_j  \s_j^z \right).
 \end{align*}

The $\delta_j\propto|\alpha|$ are now numbers, as we specified the state of the field. $\hat{H}_1$ rotates $j$-th qubit's state around the axis $\mathbf{n}$ on the Bloch sphere:
 \begin{align*}
     \bm{n} = \frac{1}{\sqrt{|\alpha^\star I_{j}^+|^2 + \lambda^2 |\delta_j|^2}}(\mathrm{Re}[\alpha^\star I_{j}^+], -\mathrm{Im}[\alpha^\star I_{j}^+] ,  -i \lambda \delta_j)
 \end{align*}
 with the magnitude of rotation $\zeta$ given by
 \begin{equation}\label{eq:mag_rot}
     \zeta = 2\lambda |\bm{n}| = 2\lambda \sqrt{|\alpha^\star I_{j}^+|^2 + \lambda^2 |\delta_j|^2}.
 \end{equation}

We note that similar expressions were found earlier in \cite{MMartinezAasenKempf}, however without the transparency assumption. Here, the single-qubit gates are performed with non-resonant interactions.

We, therefore, apply the single qubit gates by rotating around axis $\mathbf{n}$ on the Bloch sphere, by magnitude $\zeta$, over multiple trajectory cycles. We then manipulate the phase of $\alpha$ to create a sufficient set of rotation axes, thereby achieving arbitrary single-qubit gates. 

\textbf{Two-qubit gates.} Having found that we can apply arbitrary single-qubit gates, we now have the task to show that we can perform entangling gates, to ensure universal quantum computation. While the implementation of one-qubit gates presented in~\cite{Pozas-Kerstjens:2015} can be naturally applied to the non-resonant case, standard entangling techniques need to be adapted in the context of acceleration-induced transparency. One method to implement two-qubit gates on systems undergoing relativistic motion is to extract previously existing entanglement in the field state that the qubits couple to~\cite{Reznik2003,Reznik1,Valentini1991,Pozas-Kerstjens:2015,Pozas2016}. Processes of this type are commonly referred to as entanglement harvesting protocols. 

In our setup it is possible to quantify the entanglement acquired by two qubits after interacting with the field using the negativity~\cite{VidalNegativity}, a faithful entanglement quantifier for bipartite qubit systems. Using the  previously introduced interaction Hamiltonian, we find that the expressions for negativity created in our two-qubit system (see Supplementary Information). Assuming e.g. that the two qubits are initialized in their ground states, the negativity is given by: 
\begin{equation}\label{eq:negativity}
    \mathcal{N} = \max\left(\lambda^2|\mathcal{M}| - \frac{\lambda^2}{2}\langle \hat{g}^\dagger \!\hat{g}\rangle - \frac{\lambda^2}{2}|I^+|^2,0\right),
\end{equation}
where the $\mathcal{M}$ term contains information about the propagation of information through the field, and the extraction of entanglement previously present in the vacuum~\cite{ericksonNew,quantClass}. It is a somewhat complicated expression of the time integrals $I^{\pm}$, but straightforwardly interpretable in the context of entanglement harvesting, and it is not affected by the excitations of the field. The term $\langle \hat{g}^\dagger\!\hat{g}\rangle$ contains  the dependence on the field state in the mode $\bm k_{n}$ through the expected value of the operator $\hat{g}$, precisely defined in Supplementary Information. This operator, $\hat{g}$, gives us a control over the quantity of harvested entanglement. The operator itself is a linear combination of the creation and annihilation operators of the field, and its expectation value is somewhat reminiscent of the calculations of the Unruh vacuum, by use of Bogoliubov transformations.

\begin{figure}[h!]
\includegraphics[width=0.55\textwidth]{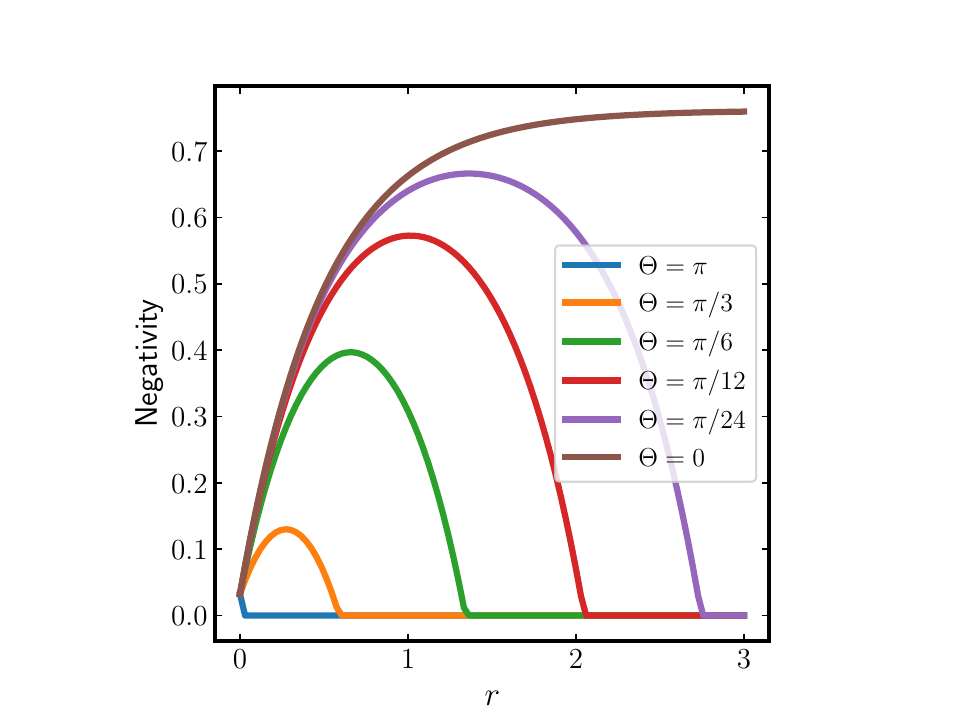}
    \caption{The negativity acquired by the two-qubit system as a function of the squeezing parameter $r$ for the optimal value of $\varphi$ for a fixed $\Omega $ and $\bm k$, when the qubit undergoes a transparent trajectory.}
    \label{fig:negativity}
\end{figure}

Since we are interested in as efficient quantum computation as possible, we looked at the expectation value $\langle \hat{g}^\dagger \!\hat{g}\rangle$ and optimized its parameters. We found that the stimulation with the coherent state of the quantum field can never give us entanglement harvesting faster than the vacuum harvesting simply because $\langle \hat{g}^\dagger \!\hat{g}\rangle \geq 0$ , and it only reduces the negativity caused by the vacuum term, proportional to $\mathcal{M}$. However, when we stimulate with a squeezed state of the field:
\begin{equation}
    \hat{S}(r,\phi)\ket{0} = e^{\frac{r}{2}\left(e^{-\ii \phi}\hat{a}^2-e^{\ii \phi}\left(\hat{a}^{\dagger}\right)^2\right)}\ket{0},
\end{equation}
with $r$ and $\phi$ squeezing parameter and angle, the negativity expression is now given by:
\begin{equation}
    \mathcal{N}=\frac{\lambda^2}{2}[\abs{\mathcal{M}}- 2\abs{I^+}^2(\cosh^2(r)-\frac{1}{2}\sinh(2r)\cos(\Theta))],
\end{equation}

where $\Theta$ is a combination of various phase factors, arising from trajectory parameters and squeezing parameters. We can now optimize this expression, obtaining \cref{fig:negativity}.

We see for example, that by tuning the phase $\Theta$, we can in principle increase the entanglement harvested by linearly increasing the squeezing parameter $r$. However, maximal $r$ realizable optically in the laboratory settings is reasonably limited to less than 2. Moreover, we can compare the entanglement created through this stimulation to the decoherence through non-resonant channel in the qubit's interaction with the field (see \cref{eq:decoherence_probability}) and it increases comparably with $e^{2r}$, therefore this is not a desirable choice of $\Theta$:
\begin{equation}
         \mathcal{P}(\bldk)_{\textup{cycle}}\propto\lambda^2 \abs{I^{+}}^2 \cosh^2(r).
   \end{equation}

Furthermore, it is interesting that for a generic choice of $\Theta$, the negativity goes to zero at a large enough $r$. However, its maximum value stays comparable to vacuum harvesting, and consequently the application of entangling gates is as inefficient as entanglement harvesting for two qubits. However, we can choose to turn off entanglement harvesting at will, by stimulating the qubit's environment with an appropriate choice of $r$-squeezed state for a given field mode, which might be an interesting phenomenon to explore for light-matter interactions in a cavity, or in a free field.

\textbf{Discussion.} Having the ability to perform all single qubit gates, plus an entangling gate between any two qubits implies a universal set of gates for universal quantum computation ~\cite{Dodd2001, Dodd2002}. We have shown the ability to perform such a set of gates. However the question of time scales arises: how efficient would an error-protected computer be?

The \emph{Eastin-Knill theorem}~\cite{EastinKnill2009} proves that no quantum error-correcting code can support a universal set of logical gates implemented transversally, provided the code can correct arbitrary local errors. This imposes a fundamental trade-off between universality and locality in  quantum error correction.

We compare this to our results. While protecting qubits from a dominant channel of decoherence we found a significant slow-down in application of entangling gates, which are needed for universal quantum computation. Conversely, we find that if we allow for the decoherence causing resonant transitions, we have a much faster application of the two-qubit gates. Therefore, we discovered a different realization of the same concept as proven in the Eastin-Knill theorem.

While Eastin-Knill theorem phrases the impossibility in terms of only being able to achieve a discrete subset of the entire (dense subset) of the group representing the universal operations, here we frame it in terms of the trade-off between speed of application of gates versus speed of loss of coherence.

Due to our framework, we can quantify the tradeoff, and compare to Eastin-Knill and similar results ~\cite{BravyiKoenig2013,Pastawski2015,Haah2011}. We will compare the time for achieving a goal negativity $\mathcal{N}^*$ of order 1 by applying consecutively an entangling gate in our setup, which we call $\tau_{gate}$, to the time for a qubit decohering with probability $p=1$, namely $\tau_{dec}$:
\begin{equation}
    \tau_{gate}\approx\frac{\mathcal{N}^*}{\mathcal{N}} T=\frac{2 T}{\lambda^2 \left(\abs{\mathcal{M}}-2\abs{I^{+}}^2 f(r,\Theta)\right)}.
\end{equation}
We will compare that time to the time $\tau_{dec}$:  
  \begin{equation}
    \tau_{dec}\approx\frac{1}{P_{\text{dec}}^{(\text{cycle})}} T=\frac{T}{\lambda^2 \abs{I^{+}}^2 \cosh^2 r} .
\end{equation}

The feasibility condition $\tau_{gate}<\tau_{dec}$ requires:
\begin{equation}
\frac{\abs{I^{+}}^2 \cosh^2 r}{\abs{\mathcal{M}}- 2\abs{I^+}^2(\cosh^2(r)-\frac{1}{2}\sinh(2r)\cos(\Theta))}<1.
\end{equation}

$\abs{\mathcal{M}}$ is a non-trivial function of the trajectory, and depends on  derivatives of the resonant integral $I^-$.

We can analyze the two regimes $\abs{\mathcal{M}} \gg \abs{I^+}^2$, $\abs{\mathcal{M}} \approx \abs{I^+}^2$, and optimize for the squeezing parameters. 

In the ideal case of excellent control over decoherence, we have that the non-resonant contribution is much smaller than the entanglement harvesting: $\abs{\mathcal{M}} \gg \abs{I^+}^2$. Then, we can achieve the ratios between times to become proportional to the ratio:
\begin{equation}
    \frac{\tau_{gate}}{\tau_{dec}}\approx \frac{\abs{I^+}^2}{\abs{\mathcal{M}}}\ll 1.
\end{equation}

For less controlled scenarios, where entanglement harvesting magnitude is comparable to non-resonant contributions, $\abs{\mathcal{M}}=\abs{I^+}^2+\delta$, $0<\delta<\abs{I^+}^2$ we can not achieve a ratio less than 1:
\begin{equation}
    \frac{\tau_{gate}}{\tau_dec}\approx \frac{\abs{I^+}^4}{\delta^4}\rightarrow \infty, \delta \rightarrow 0.
\end{equation}

Therefore, for the ratio of the time to apply gates to time of decoherence,  $\frac{\tau_{gate}}{\tau_{dec}}$, it is best to have $\abs{I^+}^2\ll\abs{\mathcal{M}}$. However, the same $\abs{I^+}^2$ controls the speed of application of gates, as can be seen in e.g. the single-qubit gate Hamiltonian. Therefore, not only the ratio of the times matters, but also the absolute value of the time to apply gates, in order to build a  useful quantum computer. Since the time to apply a single qubit gate is $\tau_{1-gate} \propto \abs{I^+}^2 $, we find a fundamental tension, similarly to the Eastin-Knill theorem: the better the protection from decoherence, the slower the application of quantum gates, and vice-versa. Here, instead of a no-go theorem, we find a physical mechanism by which protection from decoherence, slows down computation.

\textbf{Conclusion and Outlook.}

In this work, we investigated a scenario of quantum error protection: by externally controlling the classical trajectory of a qubit, we find that it is possible to shut down a dominant source of decoherence through resonant transitions. Since the strong resonant transitions are absent, we use the non-resonant terms to perform quantum gates. For these purposes, for the first time, we demonstrate transparent trajectories, i.e. ones which turn off resonant transitions, compact both in space and time, and compute a universal set of gates using non-resonant interactions to lowest orders, together with entanglement harvesting. We found that there is a fundamental trade-off between the speed of quantum computing and its protection from decoherence. This is reminiscent of the Eastin-Knill theorem, however, with a physical intuition and control over the trade-off between decoherence and computation speed. 

\begin{acknowledgements} We acknowledge helpful discussions with Asif Ayub, Achim Kempf, and Eduardo Martin-Martinez. We are grateful to Perimeter's Academic Programs, since this work started during the 2024 PSI Winter School. TRP is thankful for financial support from the Olle Engkvist Foundation (no.225-0062). This work was partially conducted while TRP was still a PhD student at the Department of Applied Mathematics at the University of Waterloo, the Institute for Quantum Computing, and the Perimeter Institute for Theoretical Physics. TRP acknowledges partial support from the Natural Sciences and Engineering Research Council of Canada (NSERC) through the Vanier Canada Graduate Scholarship. Research at Perimeter Institute is supported in part by the Government of Canada through the Department of Innovation, Science and Industry Canada and by the Province of Ontario through the Ministry of Colleges and Universities.

Perimeter Institute and the University of Waterloo are situated on the Haldimand Tract, land that was promised to the Haudenosaunee of the Six Nations of the Grand River, and is within the territory of the Neutral, Anishinaabe, and Haudenosaunee people.
\end{acknowledgements}

\bibliographystyle{apsrev4-2}
\bibliography{references}

\end{document}



\section{Time evolution and the effective Hamiltonian}\label{asec:eff_Ham}

We consider the evolution of some finite number of UDW detectors with energy gap $\Omega$. The $j\textsuperscript{th}$ detector moves on the trajectory $\bm{x}_j(\tau)$ with $\bm{x}_{j+1}(\tau) = \bm{a} + \bm{x}_j(\tau)$. $\bm{a}$ is a constant vector, therefore the UDW detectors are at a fixed distance from each other. We assume that these detectors are in a cavity with periodic boundary conditions, interacting with a scalar field. In the reference frame comoving with the detectors, the light-matter coupling Hamiltonian is the following: 
\begin{align}
    \hat{H} = &\ \frac{\text{d} t(\tau)}{\text{d} \tau}\sum_{n} |\bm k_n| \hat{a}^\dagger_n \hat{a}_n + \Omega \sum_j  \sigma^+_j \s^-_j \nonumber \\
    &+ \sum_{n, j} \lambda_n (\s^+_j + \s^-_j)(\hat{a}_n e^{i\bm{k}_n \cdot \bm{x}_j(\tau)} + \hat{a}^\dagger_n e^{-i\bm{k}_n \cdot \bm{x}_j(\tau)}), \label{2QH}
\end{align}

where $n$ labels the modes of the cavity, with the corresponding frequency $|\bm k_n|$ and wave vector $\bm{k}_n$ in the reference frame of the cavity. Here, we have used the fact the detectors share the same proper time $\tau$ and adapted the relativistic treatment of \cite{brown_detectors_2013}.

Treating the first two terms of \eqref{2QH} as free evolution, we have the corresponding interaction picture Hamiltonian:
\begin{align}
    \hat{H}_I = \sum_{n, j} \lambda_n (\s^+_j e^{i \Omega \tau} + \s^-_j e^{-i \Omega \tau})(\hat{a}_n e^{-i |\bm k_n| t(\tau) + i\bm{k}_n \cdot \bm{x}_j(\tau)} + \hat{a}^\dagger_n e^{i |\bm k_n| t(\tau)-i\bm{k}_n \cdot \bm{x}_j(\tau)}). \label{2QHI}
\end{align}

To simplify the calculations we only consider the effect of mode $n$, therefore we drop the $n$ index and define the time integrals $I_{j, s_1,s_2}$ as follows:   
\begin{align}\label{eq:Im_Ip_general}
    I_{j, s_1 s_2}(\tau) = \int_{\tau_i}^{\tau} \text{d}\tau' \ e^{i s_1 \Omega \tau' + i s_2(\omega t(\tau') -  \bm{k} \cdot \bm{x}_j(\tau'))}, \quad I_{j, +\pm}(\tau) \equiv I_{j}^\pm(\tau),
\end{align}

where $s_{1,2}$ are different sign choices $s_1, s_2 \in \{+, -\}$. Here, $\tau_i$ is the initial time at which we consider the system. We can now write:
\begin{align}
    \hat{H}_I = \lambda \frac{\text{d}}{\text{d}\tau}\sum_j \left(\s_j^+ \hat{a}^\dagger I_{j, ++} + \s_j^- \hat{a} I_{j, --} + \s_j^+ \hat{a} I_{j, +-} + \s_j^- \hat{a}^\dagger I_{j, -+}\right) \equiv \lambda \frac{\text{d}X}{\text{d}\tau}. \label{2QHISimple}
\end{align}
The operator $\hat{U}$ generated by $\hat{H}_I$, describing the time evolution between times $\tau_i$ and $\tau_f$ can be expanded at second order in $\lambda$ as follows 
\begin{align}\label{eq:UU}
    \hat{U}(\tau_i, \tau_f) = 1 + \underbrace{-i \int_{\tau_i}^{\tau_f} \mathrm{d} \tau \hat{H}_{I}(\tau)}_{\hat{U}_1} + \underbrace{-\int_{\tau_i}^{\tau_f} \mathrm{d} \tau \int_{\tau_i}^\tau \mathrm{~d} \tau' \hat{H}_{I}(\tau) \hat{H}_{I}\left(\tau'\right)}_{\hat{U}_2} + O(\lambda^3),
\end{align}
with 
\begin{align}
    \hat{U}_1 &= -i\lambda (X(\tau_f)-X(\tau_i)) = -i\lambda X(\tau_f), \label{U1}\\
    \hat{U}_2 &= -\frac{\lambda^2}{2}\int_{\tau_i}^{\tau_f} \mathrm{d} \tau \left(\frac{\text{d}X(\tau)}{\text{d}\tau} (X(\tau)-X(\tau_i)) +  X(\tau) \frac{\text{d}X(\tau)}{\text{d}\tau} +  \left[\frac{\text{d}X(\tau)}{\text{d}\tau}, X(\tau)\right]\right)  \nonumber\\
    &= -\frac{1}{2}\lambda^2 (X(\tau_f)^2 - X(\tau_i)^2) -\frac{\lambda^2}{2}\int_{\tau_i}^{\tau_f} \mathrm{d} \tau \left[\frac{\text{d}X(\tau)}{\text{d}\tau}, X(\tau)\right] \nonumber \\
    &= \frac{1}{2}\hat{U}_1^2 -\frac{\lambda^2}{2}\int_{\tau_i}^{\tau_f} \mathrm{d} \tau \left[\frac{\text{d}X(\tau)}{\text{d}\tau}, X(\tau)\right] \quad \text{(using $X(\tau_i) = 0$)} \label{U2}.
\end{align}
Again, the calculation is simplified by denoting $\hat{a} = \hat{a}^-$ and $\hat{a}^\dagger = \hat{a}^+$. With these symbols, the commutator obtained in \eqref{U2} may be explicitly expressed as: 
\begin{align}\label{eq:U2_and_U_prime}
    \left[\sum_{j, s_1 s_2} \s_j^{s_1} \hat{a}^{s_2} \frac{\text{d} I_{j, s_1 s_2}}{\text{d}\tau}, \sum_{k, s_3 s_4} \s_k^{s_3} \hat{a}^{s_4} I_{k, s_3 s_4}\right] = \sum_{j, s_1 s_2} \sum_{k, s_3 s_4} \frac{\text{d} I_{j, s_1 s_2}}{\text{d}\tau}  I_{k, s_3 s_4}  \left[\s_j^{s_1} \hat{a}^{s_2} ,  \s_k^{s_3} \hat{a}^{s_4} \right].
\end{align}
For the operator part of the above commutator, i.e. $\left[\s_j^{s_1} \hat{a}^{s_2} ,  \s_k^{s_3} \hat{a}^{s_4} \right]$, we get: 
\begin{align*}
    \left[\s_j^{s_1} \hat{a}^{s_2} ,  \s_k^{s_3} \hat{a}^{s_4} \right] &= \s_j^{s_1} \left[ \hat{a}^{s_2} ,  \s_k^{s_3} \hat{a}^{s_4} \right] + \left[\s_j^{s_1},  \s_k^{s_3} \hat{a}^{s_4} \right] \hat{a}^{s_2}\\
    &= \s_j^{s_1}\s_k^{s_3} \left[ \hat{a}^{s_2},   \hat{a}^{s_4} \right]
    + \left[\s_j^{s_1},  \s_k^{s_3} \right]  \hat{a}^{s_4} \hat{a}^{s_2} \\
    &=\s_j^{s_1}\s_k^{s_3} \epsilon_{s_2 s_4}
    - \delta_{jk} \s_j^z \epsilon_{s_1 s_3} \hat{a}^{s_4} \hat{a}^{s_2}.
\end{align*}
Note that $\left[ \hat{a}^{-},   \hat{a}^{+} \right] = -\left[ \hat{a}^{+},   \hat{a}^{-} \right] = 1$ leads to $\epsilon_{s_2 s_4}$ being the anti-symmetric symbol with $\epsilon_{-+} = 1$. Similarly, $\left[\s_j^{+}, \s_j^{-} \right] = -\left[\s_j^{-}, \s_j^{+} \right] = \s_j^z$.


With this result, we have 
\begin{align}
    \hat{U}_1 &= -\ii \lambda \sum_j \left(\s_j^+ \hat{a}^\dagger I_{j, ++} + \s_j^+ \hat{a} I_{j, +-} + \text{h.c.}\right),\\
    \hat{U}_2 &= -\frac{\lambda^2}{2} \sum_{j, k} \left(\s_j^+ \hat{a}^\dagger I_{j, ++} + \s_j^+ \hat{a} I_{j, +-}+ \text{h.c.}\right)\left(\s_k^+ \hat{a}^\dagger I_{k, ++} + \s_k^+ \hat{a} I_{k, +-} + \text{h.c.}\right)\nonumber\\
    &\quad +\lambda^2\sum_{j< k} \sum_{s_1, s_3} (A_{s_1 s_3}^{(jk)}+A_{s_1 s_3}^{(kj)}) \s_j^{s_1}\s_k^{s_3} + \lambda^2\sum_j \hat{\delta}_j  \s_j^z + \lambda^2\sum_{j} \frac{1}{2}( A_{-+}^{(jj)} + A_{+-}^{(jj)})\mathbb{I},
\end{align}
with 
\begin{align}\label{eq:A_B_coeff}
    &A_{s_1 s_3}^{(jk)} = -\frac{1}{2}\sum_{s_2 s_4}  \int_{\tau_i}^{\tau_f} \mathrm{d} \tau \frac{\text{d} I_{j, s_1 s_2}}{\text{d}\tau}  I_{k, s_3 s_4} \epsilon_{s_2 s_4},\quad B_{s_2 s_4}^{(j)} = \frac{1}{2}\sum_{s_1 s_3}  \int_{\tau_i}^{\tau_f} \mathrm{d} \tau \frac{\text{d} I_{j, s_1 s_2}}{\text{d}\tau}  I_{j, s_3 s_4} \epsilon_{s_1 s_3},\\\label{eq:delta_j}
    &\hat{\delta}_j = \sum_{s_2, s_4} B_{s_2 s_4}^{(j)} \hat{a}^{s_4} \hat{a}^{s_2} - \frac{1}{2}A_{s_2 s_4}^{(jj)} \epsilon_{s_2 s_4}.
\end{align}
Note that the last term in $\hat{U}_2$ is a constant multiple of identity and will be ignored in what follows. 

The effective Hamiltonian $\hat{H}_{\rm eff} = \hat{H}_1 + \hat{H}_2$ corresponding to $\hat{U}(0, T)$ is given by 
\begin{align}\label{eq:H1}
     &\hat{H}_1 =  \frac{\lambda}{T} \sum_j \left(\s_j^+ \hat{a}^\dagger I_{j, ++} + \s_j^+ \hat{a} I_{j, +-} + \text{h.c.}\right) +\ii\frac{\lambda^2}{T}\sum_j \hat{\delta}_j  \s_j^z,\\\label{eq:H2}
     &\hat{H}_2 = \ii\frac{\lambda^2}{T}\sum_{j> k} \sum_{s_1, s_3} (A_{s_1 s_3}^{(jk)}+A_{s_1 s_3}^{(kj)}) \s_j^{s_1}\s_k^{s_3}, 
 \end{align}
so that
 \begin{align*}
     \hat{U}(0, T) = 1 + \hat{U}_1 + \hat{U}_2 + O(\lambda^3) \sim 1 - \ii T (\hat{H}_1 + \hat{H}_2) -\frac {T^2}{2} (\hat{H}_1+\hat{H}_2)^2 ,
 \end{align*}
 where $\sim$ represents equality up to $O(\lambda^2)$. While $\hat{H}_1$ is used to implement single detector gates for high stimulation with a cavity coherent state, $\hat{H}_2$ provides cavity independent interaction between the detector through the matrix $A_{s_1 s_3}$. Interaction between the cavity and detectors induces a detuning of $\Omega$ parameterized by $\hat{\delta}_j$ which depends on the  state of the cavity.

\section{Transparent trajectories and periodicity constraint}
\subsection{Setting up the trajectory}
The trajectory in the scheme we propose is a repetition of identical cycles composed of inertial segments labeled with $\rm s$ happening between $\tau_{\rm s}$ and $\tau_{\rm s+1}$. More precisely, the position four-vector $x^\mu$ is defined piece-wise through the four-velocity $v^\mu$ in each segment: 
\begin{align*}
    v^\mu = \frac{\text{d}x^\mu}{\text{d}\tau} = 
    \begin{cases}
        \gamma_{\rm s} (1, v_{\rm s}\hat{x}), \quad   \tau_{\rm s} \leq \tau < \tau_{\rm s + 1} \quad \text{segment (s)},
    \end{cases}
\end{align*}
with $\gamma_{\rm s} = (1-v_{\rm s}^2)^{-1/2}$. The associated position four-vector in segment $(\rm s)$ between times $\tau_{\rm s}$ and $\tau_{\rm s+1}$ is of the form $\bm{x}(\tau) = \gamma_{\rm s} v_{\rm s} (\tau - \tau_{\rm s}) \hat{x} + \bm{x}(\tau_{\rm s})$ with $\bm{x}(\tau_{\rm s}) \equiv x_{\rm s} \hat{x}$ and $t(\tau) = \gamma_s (\tau-\tau_s) + t(\tau_{\rm s})$ with $t(\tau_{\rm s}) \equiv t_{\rm s}$. 

These integrals are first computed for a single segment. We have:
\begin{align*}
    I^{-}&(v_{\rm s}, \tau_{\rm s}, \tau_{\rm s+1}) = \int_{\tau_{\rm s}}^{\tau_{\rm s + 1}} \text{d}\tau \ \exp(i\Omega \tau - i k (t(\tau) - x(\tau)))\\
    &= \int_{\tau_{\rm s}}^{\tau_{\rm s + 1}} \text{d}\tau \ \exp(i\Omega \tau - i k (\gamma_s (\tau-\tau_s) + t_{\rm s} - \gamma_{\rm s} v_{\rm s} (\tau - \tau_{\rm s}) - x_{\rm s})))\\
    &= \exp(-i k (t_{\rm s} + \gamma_{\rm s} (v_{\rm s}-1) \tau_{\rm s} - x_{\rm s}) + i\frac{(\Omega-k_{\rm s})(\tau_s + \tau_{\rm s+1})}{2} ) \frac{\sin(\frac{(\Omega - k_{\rm s}) (\tau_{\rm s+1} - \tau_{\rm s})}{2})}{\frac{\Omega - k_{\rm s}}{2}}.
\end{align*}
We are looking for trajectories for which $I^{-} = 0$ at times $\tau_{\rm r}$, which constitutes one cycle. Each such cycle consists of $r \in \mathbb{N}$ such segments in which case, the transparency condition reads: 
\begin{align}\label{eq:transparency_condition}
    0 = \sum_{\rm s=0}^{\rm r-1} I^{-}(v_{\rm s}, \tau_{\rm s}, \tau_{\rm s+1}). 
\end{align}
This means that all the inertial segments in each cycle destructively interfere with each other so that the $I^-$ integral at the end of the cycle vanishes. Thus, at the end of each cycle, the resonant terms will not contribute to the qubit-light interaction, effectively shielding the qubits from decoherence through the primary decoherence channel.

\begin{figure}
    \hspace*{1.5cm}\includegraphics[width=0.38\linewidth]{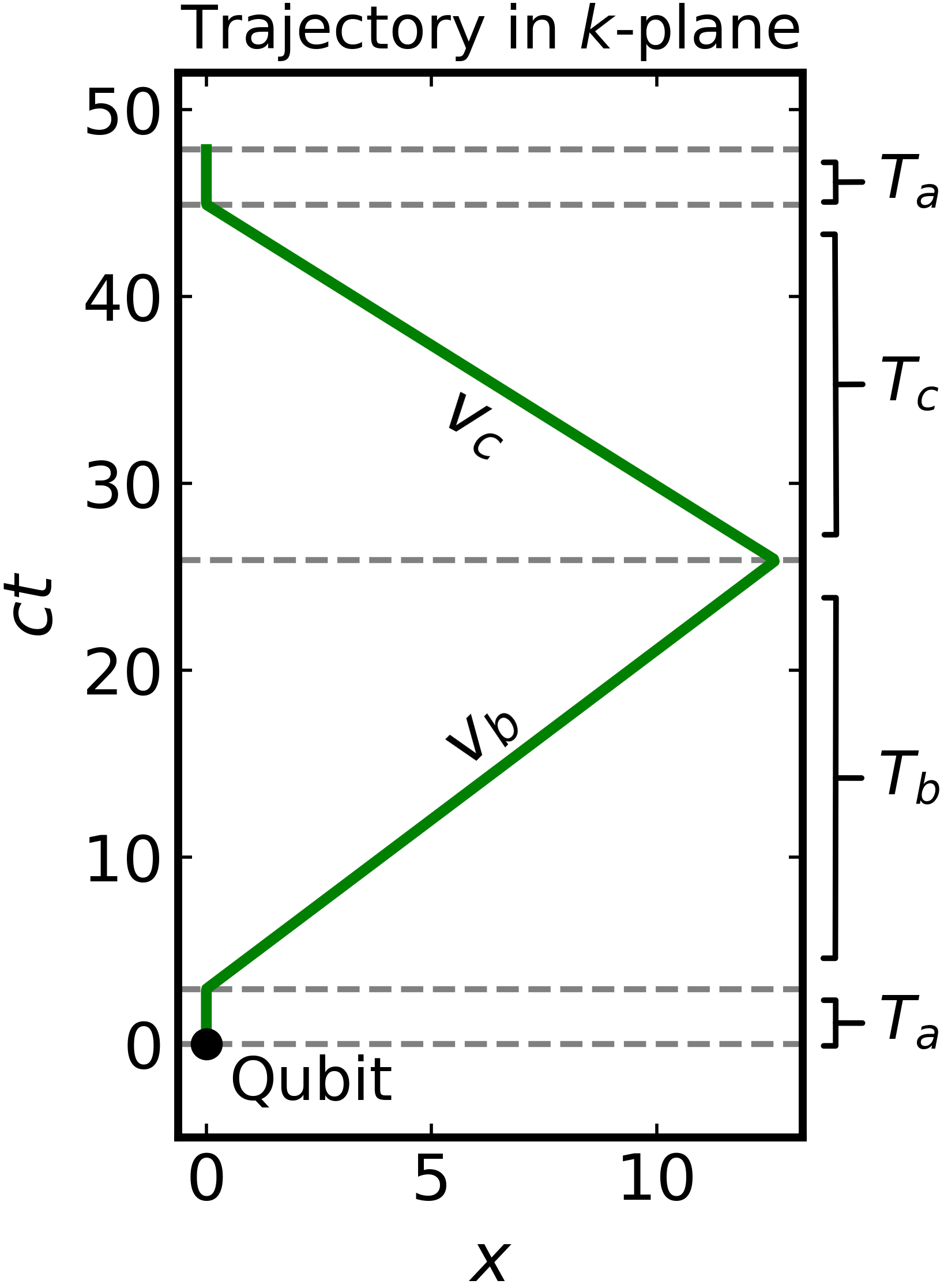}
    \caption{Plot of one cycle, composed of 4 inertial segments: first, standing still, second and third moving at different velocities, and fourth standing still.}
    \label{fig:trajectories}
\end{figure}

For our initial analysis, we consider a trajectory with 
$r=4$ segments, where the first and last segments are at rest, while the second and third segments have velocities $v_b$ and $v_c$ respectively. Let the associated coordinate times with each segment be given by $T_a$, $T_b$, $T_c$ and then $T_a$ again. The piecewise trajectory is the following:
\begin{align*}
    v^\mu = \frac{\text{d}x^\mu}{\text{d}\tau} = 
    \begin{cases}
        (1, \bm{0}), \quad   0 \leq \tau < \tau_1 \quad \text{(a)}\\
        \gamma_b (1,  v_b \hat{x}), \quad \tau_1 \leq \tau < \tau_2\quad\text{(b)}\\
        \gamma_c (1, -v_c \hat{x}), \quad \tau_2 \leq \tau < \tau_3\quad \text{(c)}\\
        (1, \bm{0}), \quad \tau_3 \leq \tau < \tau_4 \quad \text{(d)},\\
    \end{cases}
\end{align*}
with proper times $\tau_0 = 0,\ \tau_1 = \tau_a,\ \tau_2 = \tau_a+\tau_b,\ \tau_3 = \tau_a + \tau_b+\tau_c$ and $\tau_4 = 2\tau_a + \tau_b+\tau_c$. This trajectory has been depicted in Fig.~\ref{fig:trajectories}.

For the qubit to return to the same position (i.e. the origin) at time $\tau_3$, we have the following (trajectory) constraint:
\begin{equation}\label{eq:traj_constraint}
    v_b T_b = v_c T_c.
\end{equation}
\subsection{Periodicity constraint}
The trajectory design we considered ensures that the qubit’s position, $x(\tau)$, is periodic after each cycle, i.e., its position at corresponding times in successive cycles remains the same. This naturally raises the question of whether one can also obtain a Hamiltonian that is periodic within each cycle. Enforcing such a periodicity constraint guarantees that the phase at the beginning and end of each cycle match, preventing any spurious phase accumulation in the calculation of both one-qubit and two-qubit gates. 

A periodic Hamiltonian corresponds to the periodicity of $I^\pm$ integrals given in Eq.~\eqref{eq:Im_Ip_general}. Since, $x(\tau)$ is periodic by design, we require the remaining terms in the exponential of $I^\pm$ to be an integer multiples of $2\pi$. Thus, in the interaction picture, this periodicity constraint translates to: 
\begin{equation}
 \Omega \tau + k t(\tau) = 2\pi m_+ \quad \& \quad \Omega \tau - k t(\tau) = 2\pi m_-,
\end{equation}
for $m_+, m_- \in \mathbb{Z}$. Using $\tau=\tau_4$ (time at the end of a cycle), the above equations look like: 
\begin{equation}\label{eq:Floquet_constraint}
    2T_a + \frac{T_b}{\gamma_b}+\frac{T_c}{\gamma_c} = \frac{\pi n}{\Omega}=n_o \quad \&\quad 2T_a +T_b+T_c=\frac{\pi m}{k}=m_k,
\end{equation}
where $n=m_++m_-$ and $m=m_+-m_-$.
Solving the above two constraints using Eq.~\eqref{eq:traj_constraint}, we get 
\begin{equation}\label{eq:Tb}
    T_b = \frac{m_k-n_o}{1-\frac{1}{\gamma_b}+\frac{v_b}{v_c}(1-\frac{1}{\gamma_c})},
\end{equation}
and 
\begin{equation}\label{eq:gammac}
    \gamma_b = \frac{2p(1-p)(1-\gamma_c)+\gamma_c}{2p(1-p)(1-\gamma_c)-1},
\end{equation}
where 
$$p = \frac{m_k-2T_a}{m_k-n_o}$$

$$p=\frac{T_b+T_c}{T_b(1-\frac{1}{\gamma_b})+T_c(1-\frac{1}{\gamma_c})},$$

$$p=\frac{T_b+T_c}{T_b+T_c-\frac{T_b}{\gamma_b}-\frac{T_c}{\gamma_c}}.$$

($p$ is measuring the ratio between coordinate time, and the difference between coordinate and proper time.)

Given this periodicity condition, we need to find the parameters for which $I^-(\Omega,k) = 0$ at the end of each cycle. Thus, in our quest for transparent trajectories, we must perform a numerical search through a parameter space, following an approach similar to that of previous work \cite{barbara}.

In the code: we fixed: $k, \Omega$, $m, n$ and vary: $T_a, v_c$. We find that sometimes $\gamma_b$ is negative, which is unphysical. To fix that, we use a condition on relationship between $T_a$ and $\gamma_c$, which is telling us that we can't have $p \rightarrow 1$ unless $\gamma_b$ and $\gamma_c$ are very large:

$$ 2 \frac{m_k -2 T_a}{m_k - n_o}\frac{n_o-2 T_a}{m_k - n_o}>\frac{1}{\gamma_c-1},$$
\begin{equation}\label{eq:gamma2_constraint}
    \Rightarrow \gamma_c >\frac{(m_k-n_o)^2}{2(m_k-2T_a)(n_o-2T_a)} + 1.
\end{equation}
In terms of the velocity $v_c$, this constraint looks like:
$$v_c > \sqrt{1-\frac{1}{{\gamma'}^2_c}},$$
$$\gamma'_c = \frac{(m_k-n_o)^2}{2(m_k-2T_a)(n_o-2T_a)} + 1.$$
In summary, we conduct a numerical search to identify the optimal parameters that satisfy the trajectory constraint \eqref{eq:traj_constraint}, the periodicity constraint \eqref{eq:Floquet_constraint}, the non-negativity condition on $\gamma_b$ \eqref{eq:gamma2_constraint}, along with the transparency condition \eqref{eq:transparency_condition}. Since, $I^-$ is a complex function, we determine its true zeros by separately locating the zeros of its real and imaginary parts and finding their intersection points.  
 
\subsection{Transparency with periodicity}


\begin{figure}[H]
    \centering
    \includegraphics[width=\linewidth]{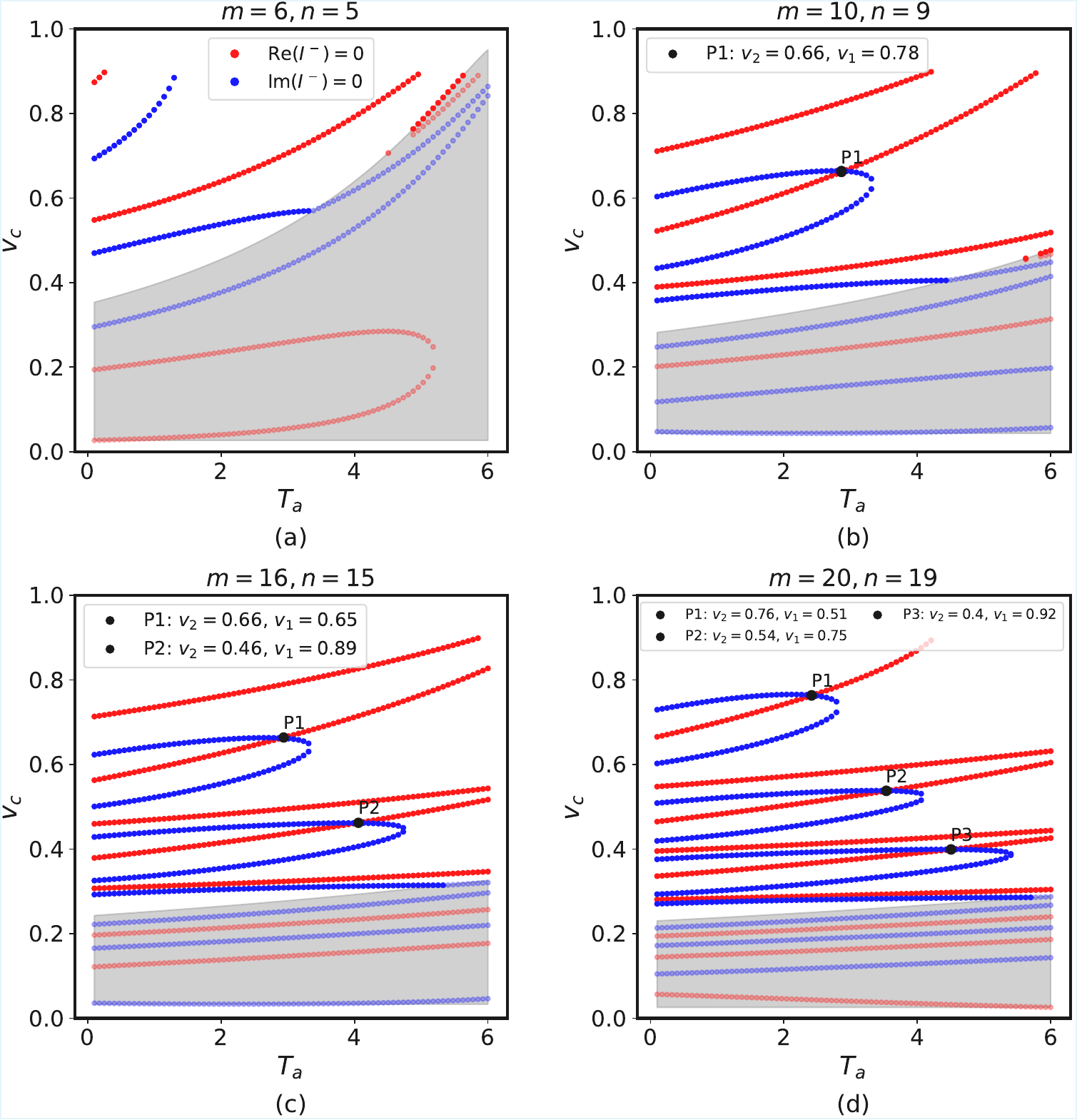}
    \caption{The zeroes of real (red) and imaginary (blue) parts of the $I^-$ integral are shown in the parameter space of $v_c$ which is the velocity during the second half of the trajectory and the initial inertial time $T_a$ for different values of $m$ and $n$ after imposing the periodicity constraint \eqref{eq:Floquet_constraint}. We have used $k=1$ and $\Omega=1.2$ for the plots. The gray area depicts the region where the value of $\gamma_b$ is negative and hence is unphysical. The intersection points (black) are the true zeroes of $I^-$ respecting the periodicity condition.}
    \label{fig:Floquet_diff_m_n}
\end{figure}

\begin{figure}
    \centering
    \includegraphics[width=\linewidth]{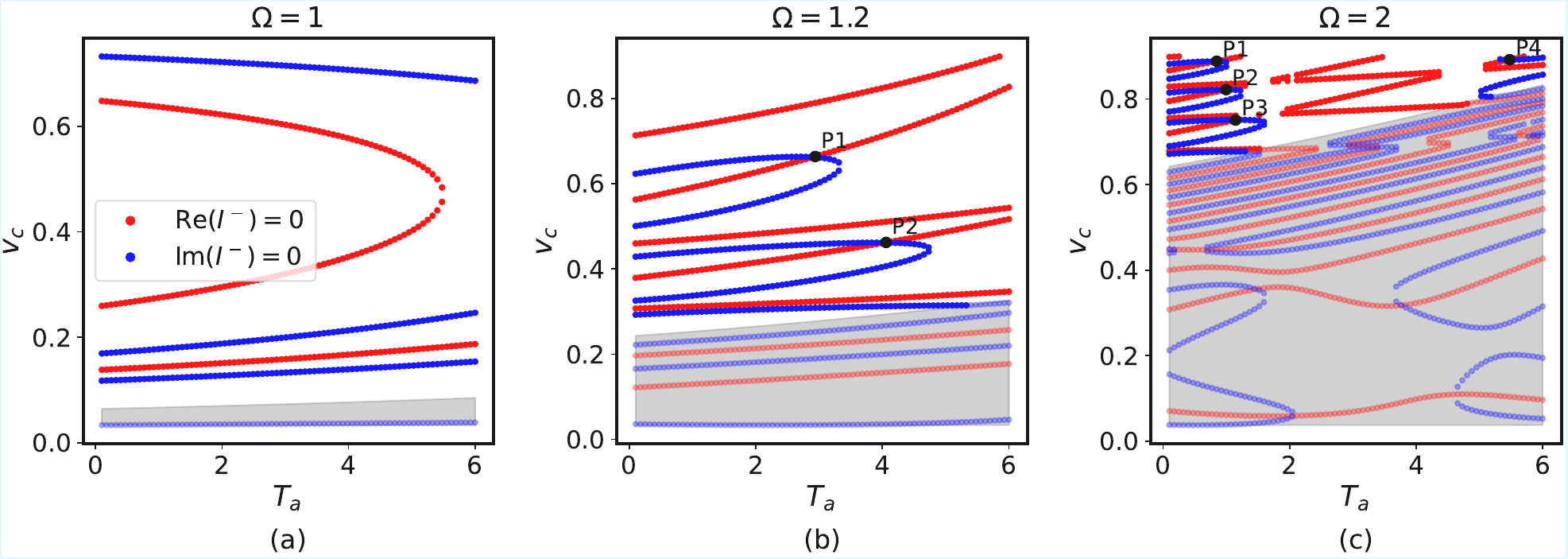}
    \caption{The zeroes of the real and imaginary parts of $I^-$ are shown in the parameter space of $v_c$ and $T_a$ for different values of the qubit energy gap $\Omega$. For each of the plots, $m=16,n=15$ and $k=1$.}
    \label{fig:Floquet_diff_Omega}
\end{figure}

Figure~\ref{fig:Floquet_diff_m_n} shows the parameter values for which we have a zero of the $I^-$ integral (intersection of blue and red curves) after imposing the periodicity constraint \eqref{eq:Floquet_constraint}. The values of $m$ and $n$ are chosen such that their difference is 1, ensuring that the velocity $v_c$ is minimum possible for this difference (see Eq.~\eqref{eq:gamma2_constraint}). This is because excessively high velocities for the qubit trajectory would be practically infeasible. From Fig.~\ref{fig:Floquet_diff_m_n}, we observe that increasing the values of $m$ and $n$ leads to greater number of intersection points and hence more solutions. We also see that the validity region for $v_c$ (evaluated from Eq.~\eqref{eq:gamma2_constraint}) becomes larger as we go to higher $m$ and $n$. However, as new solutions appear at lower $v_c$ values, the corresponding values of $v_b$ are significantly higher. This can be seen from the intersection points in Fig.~\ref{fig:Floquet_diff_m_n}(d). Consequently, for the entanglement analysis in Sec.~\ref{asec:entanglement2}, the parameter values have been chosen corresponding to point P1 in Fig.~\ref{fig:Floquet_diff_m_n}(c) where the velocity values are $v_b=0.65$ in the first half of the cycle and $v_c=0.66$ in the second half for both the qubits-- a well-balanced choice of $v_b$ and $v_c$ values considering all relevant factors.

Figure~\ref{fig:Floquet_diff_Omega} shows dependence of zeroes of $I^-$ integral on the qubit energy level gap $\Omega$. We note that for $\Omega=1=k$, there are no intersection points; in fact this is observed for any $\Omega=k$. We also see that as $\Omega$ increases, the region of validity of $v_c$ decreases. This makes another case for choosing Fig.~\ref{fig:Floquet_diff_m_n}(c) (which same as Fig.~\ref{fig:Floquet_diff_Omega}(b)) for future analysis. 

\subsection{The trajectory that has both transparency and periodicity}

\begin{figure}
    \centering
    \includegraphics[width=\linewidth]{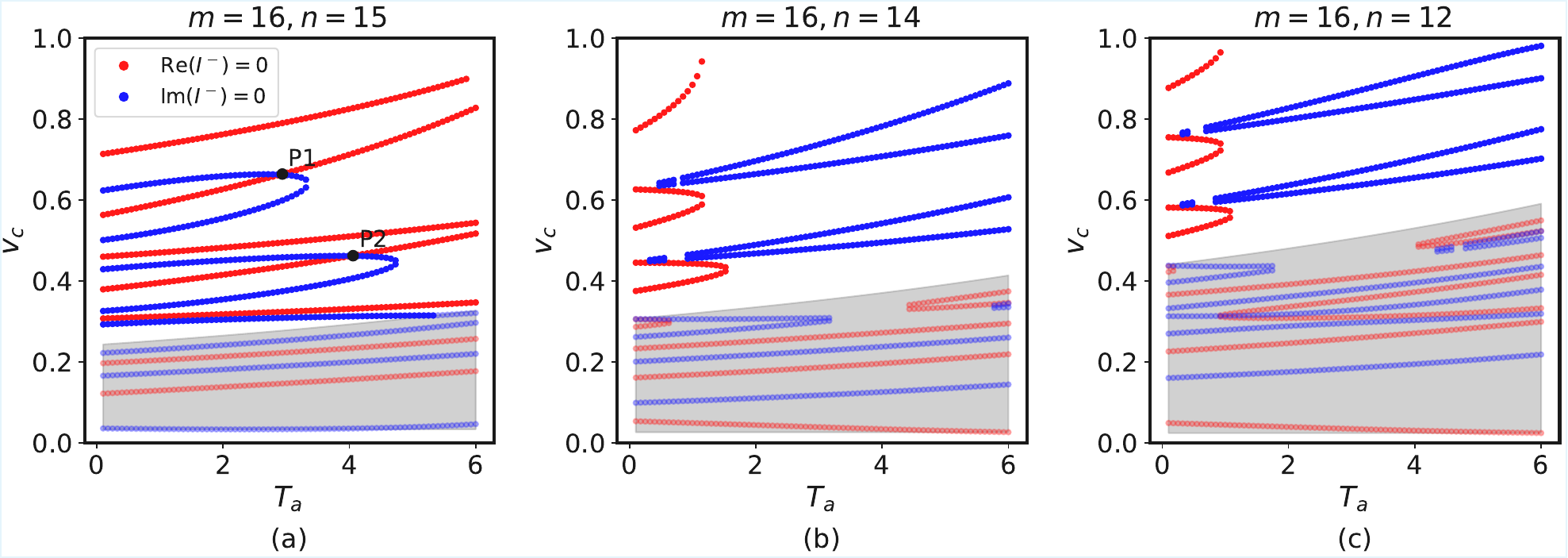}
    \caption{The zeroes of real and imaginary parts of $I^-$ shown in the parameter space of $v_c$ and $T_a$ for different values of $m$ and $n$. For each plot, $\Omega=1.2$ and $k=1$. Two intersections between the red and blue curves are observed in panel (a), whereas no such intersections appear in panels (b) and (c).}
    \label{fig:compare_diff_m_n}
\end{figure}

Both Fig.~\ref{fig:Floquet_diff_m_n} and Fig.~\ref{fig:Floquet_diff_Omega} correspond to the case where $m - n = 1$. The periodicity condition~\eqref{eq:Floquet_constraint} is expressed in terms of the parameters $m_+$ and $m_-$, given by
$$m_+ = \frac{n+m}{2},\hspace{1.5cm}m_-=\frac{n-m}{2}.$$
Clearly, for the choice of $m$ and $n$ used in Figs.~\ref{fig:Floquet_diff_m_n} and \ref{fig:Floquet_diff_Omega}, $m_+$ and $m_-$ are not integers (rather half-integers). Consequently, the periodicity condition \eqref{eq:Floquet_constraint} is not satisfied. However, the transparency condition \eqref{eq:transparency_condition} is only satisfied for $m-n=1$. This is illustrated in Fig.~\ref{fig:compare_diff_m_n}, where the zeros of the $I^-$ integral appear only in panel~(a), corresponding to $m = 16$, $n = 15$. For $m - n > 1$, our scan over the parameter space reveals no intersections between the real and imaginary parts of the $I^- = 0$ curves.

To construct a trajectory that satisfies both the transparency and periodicity conditions, we begin by noting that for $m - n = 1$, the phase accumulated at the end of a single cycle is an integer multiple of $\pi$, since both $m_+$ and $m_-$ are half-integers. If we consider two such trajectories back-to-back, the total phase accumulated over the two cycles becomes $2\pi$, thereby ensuring exact periodicity.

We therefore consider a full trajectory composed of two intervals, each resembling the transparent trajectory shown in Fig.~\ref{fig:trajectories}. However, to avoid trivial cancellation of the interaction, the parameters of the two intervals must differ.  If $I_1^\pm$ and $I_2^\pm$ are the contributions from the first and second intervals respectively, then each satisfies the transparency condition individually ($I_1^- \approx 0 \approx I_2^-$), while allowing for a nonzero amplitude via $I^+ = I_1^+ - I_2^+$. This ensures that the qubit interaction obeys the transparency condition while maintaining overall periodicity. 

\begin{figure}
    \centering
    \includegraphics[width=0.4\linewidth]{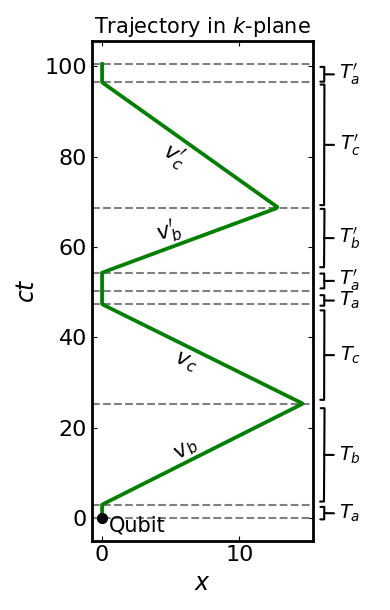}
    \caption{Plot of the trajectory that satisfies both transparency \eqref{eq:transparency_condition} and periodicity \eqref{eq:Floquet_constraint} constraints. The parameters used in the first and second intervals are the ones corresponding to the intersection points P1 and P2 from Fig.~\ref{fig:compare_diff_m_n}(a) respectively; $m=16,n=15, \Omega =1.2,k=1, T_a=2.94,T'_a=4.06, v_c = 0.66,v'_c=0.46$. The rest of the parameters can be obtained from Eqs.~\eqref{eq:traj_constraint},\eqref{eq:Tb},\eqref{eq:gammac}.}
    \label{fig:final_trajectory}
\end{figure}

Figure~\ref{fig:final_trajectory} shows such a qubit trajectory, where the first interval uses parameters corresponding to the intersection point P1 in Fig.~\ref{fig:compare_diff_m_n}(a), and the second interval is constructed using parameters from the intersection point P2. Hence, each interval individually satisfies the transparency condition, and the total phase accumulated over the full cycle is zero. We will be using the parameters of this trajectory for implementing our gates, particularly, the two-qubit gate.

\section{Implementing gate operations}
\subsection{Single-qubit gate}
After setting up the trajectories that ensure $I_{j, -+} = I_{j, +-} = 0$ for the j\textsuperscript{th} detector, lets go back to the effective Hamiltonian (Eqs.~\eqref{eq:H1} and \eqref{eq:H2}) from Sec.~\ref{asec:eff_Ham}. In this setting, the single detector contribution to the effective Hamiltonian becomes 
 \begin{align*}
     \hat{H}_1 &=  \frac{\lambda}{T} \sum_j  \left(\s_j^+ \hat{a}^\dagger I_{j, ++} + \s_j^- \hat{a} I_{j, --} -i \lambda \delta_j  \s_j^z \right) \\
     &= \frac{\lambda}{T} \sum_j  \left(\frac{1}{2}(\hat{\sigma}^x_j + i \hat{\sigma}^y_j) \hat{a}^\dagger I_{j, ++} + \frac{1}{2}(\hat{\sigma}^x_j - i \hat{\sigma}^y_j) \hat{a} I_{j, --} -i \lambda \delta_j  \s_j^z \right)\\
     &= \frac{\lambda}{T} \sum_j  \left(\frac{1}{2}(\hat{a}^\dagger I_{j, ++} + \hat{a} I_{j, --})\hat{\sigma}^x_j - \frac{1}{2i}(\hat{a}^\dagger I_{j, ++} - \hat{a} I_{j, --}) \hat{\sigma}^y_j -i \lambda \delta_j  \s_j^z \right)
 \end{align*}
 where we use $\hat{\sigma}^+_j = (\hat{\sigma}^x_j + i \hat{\sigma}^y_j)/2$.

 To use $\hat{H}_1$ to perform single detector operations, the cavity field is set to be a coherent state $\ket{\alpha}$ ($\hat{a}\ket{\alpha} = \alpha \ket{\alpha}$). Since the non-resonant terms need high stimulation to manifest meaningfully \cite{barbara}, we take the strong drive approximation and neglect the fluctuation of the electromagnetic field in the cavity. This allows~\cite{blaisCircuitQuantumElectrodynamics2021} to perform the substitution $\hat{a} \to \alpha$ and $\hat{a}^\dagger \to \alpha^\star$  and get
 \begin{align*}
     \hat{H}_1 
     &= \frac{\lambda}{T} \sum_j  \left(\mathrm{Re}[\alpha^\star I_{j, ++}]\hat{\sigma}^x_j - \mathrm{Im}[\alpha^\star I_{j, ++}] \hat{\sigma}^y_j -i \lambda \delta_j  \s_j^z \right)
 \end{align*}
 where $\delta_j$ are now numbers dependent on $\alpha$.
 
 The action of this Hamiltonian on detector $j$ is to rotate its state around the Bloch sphere axis given by the unit vector
 \begin{align*}
     \bm{n} = \frac{1}{\sqrt{|\alpha^\star I_{j, ++}|^2 + \lambda^2 |\delta_j|^2}}(\mathrm{Re}[\alpha^\star I_{j, ++}], -\mathrm{Im}[\alpha^\star I_{j, ++}] ,  -i \lambda \delta_j)
 \end{align*}
 with the magnitude of rotation $\zeta$ given by
 \begin{equation}\label{eq:mag_rot}
     \zeta = 2\lambda |\bm{n}| = 2\lambda \sqrt{|\alpha^\star I_{j, ++}|^2 + \lambda^2 |\delta_j|^2}
 \end{equation}
Note that $\delta_j$ term is a second order correction and contains contribution from stimulation via the coherent state and also vacuum contribution (see Eq.\eqref{eq:delta_j}). We can suppress vaccuum contributions by choosing $\alpha$ to be very high.
 
In our setup, for trajectories satisfying the periodicity constraint, the rotation angle scales linearly with the number of cycles $N$, as both $I_{j,++}$ and $\delta_j$ are proportional to $N$. In practice $\alpha$ is selected so that the desired operation is completed in an integer number of computation steps with duration $T$. By varying the phase of $\alpha$, a sufficient set of rotation axes can be generated to implement arbitrary single-qubit gates.

\subsection{Two-qubit entangling gate}

Now, we calculate the effect of $\hat{U}$ (given in Eqs.~\eqref{eq:UU},\eqref{U1} and \eqref{U2}) up to $O(\lambda^2)$ on the state $\hat{\rho} 
= \hat{\rho}_i \otimes \ket{\phi} \bra{\phi}$, where $\hat{\rho}_i = \ket{00} \bra{00}$ describes detectors initialized in their ground states and $\ket{\phi}$ gives the initial state of the field.
To model our ignorance of the field state, we trace it out at the end. With $\hat{U}' = \hat{U}_2-\hat{U}_1^2/2$, the final state of the detectors $\hat{\rho}_f$ has the expression: 
\begin{align*}
    \hat{\rho}_f &= \text{Tr}_\phi\left[(1+\hat{U}_1+\hat{U}_2) \hat{\rho} (1+\hat{U}_1+\hat{U}_2)^\dagger\right]\\
    &=\hat{\rho}_i + \text{Tr}_\phi\left[ \hat{U}_1 \hat{\rho} + \hat{\rho} \hat{U}_1^\dagger + \hat{U}_1 \hat{\rho} \hat{U}_1^\dagger + \frac{1}{2}\hat{U}_1^2 \hat{\rho} + \hat{\rho} \frac{1}{2}(\hat{U}_1^\dagger)^2 + \hat{U}' \hat{\rho} + \hat{\rho}\hat{U}'{}^\dagger \right].  
\end{align*}
We expand this result starting with the term $\text{Tr}_\phi [\hat{U}' \hat{\rho}]$. From Eq.~\eqref{eq:U2_and_U_prime}, we have
\begin{align}
    &\text{Tr}_\phi [\hat{U}' \hat{\rho}] = -\frac{\lambda^2}{2}\int_{\tau_i}^{\tau_f} \mathrm{d} \tau \sum_{n} \bra{n}\left(\sum_{j, s_1 s_2} \sum_{k, s_3 s_4} \frac{\text{d} I_{j, s_1 s_2}}{\text{d}\tau}  I_{k, s_3 s_4}  \left(\s_j^{s_1}\s_k^{s_3} \epsilon_{s_2 s_4}
    - \delta_{jk} \s_j^z \epsilon_{s_1 s_3} \hat{a}^{s_4} \hat{a}^{s_2}\right)  \hat{\rho}\right)\ket{n} \nonumber\\
    &= -\frac{\lambda^2}{2}\int_{\tau_i}^{\tau_f} \mathrm{d} \tau \sum_{n} \bra{n}\left(\sum_{j, s_1 s_2} \sum_{k, s_3 s_4} \frac{\text{d} I_{j, s_1 s_2}}{\text{d}\tau}  I_{k, s_3 s_4}  \left(\s_j^{s_1}\s_k^{s_3} \epsilon_{s_2 s_4}
    - \delta_{jk} \s_j^z \epsilon_{s_1 s_3} \hat{a}^{s_4} \hat{a}^{s_2}\right) \ket{\phi}\bra{\phi}  \right)\ket{n} \ket{00}\bra{00} \nonumber\\
    &= -\frac{\lambda^2}{2}\int_{\tau_i}^{\tau_f} \mathrm{d} \tau \sum_{n} \braket{\phi}{n}\bra{n}\left(\sum_{j, s_1 s_2} \sum_{k, s_3 s_4} \frac{\text{d} I_{j, s_1 s_2}}{\text{d}\tau}  I_{k, s_3 s_4}  \left(\s_j^{s_1}\s_k^{s_3} \epsilon_{s_2 s_4}
    - \delta_{jk} \s_j^z \epsilon_{s_1 s_3} \hat{a}^{s_4} \hat{a}^{s_2}\right) \ket{\phi}  \right) \ket{00}\bra{00} \nonumber\\
    &=-\frac{\lambda^2}{2}\int_{\tau_i}^{\tau_f} \mathrm{d} \tau \left(\sum_{j, s_1 s_2} \sum_{k, s_3 s_4} \frac{\text{d} I_{j, s_1 s_2}}{\text{d}\tau}  I_{k, s_3 s_4}  \left(\s_j^{s_1}\s_k^{s_3} \epsilon_{s_2 s_4}
    - \delta_{jk} \s_j^z \epsilon_{s_1 s_3} \langle\hat{a}^{s_4}\hat{a}^{s_2}\rangle\right) \right) \ket{00}\bra{00}, \nonumber
\end{align}
where $\langle\cdots \rangle = \bra{\phi} \cdots \ket{\phi}$.

Since $\s^s$ matrices only act on one side of $\ket{00}\bra{00}$ the only non-zero diagonal element in $\text{Tr}_\phi [\hat{U}'\hat{\rho} +  \hat{\rho}(\hat{U}')^\dagger]$ has to be proportional to $\ket{00}\bra{00}$. Then, because $\hat{U}'$ is the commutator of Hermitian matrices, it is anti-Hermitian and the non-trivial diagonal element has to be: 
\begin{align*}
   & \bra{00}\text{Tr}_\phi [\hat{U}'\hat{\rho} +  \hat{\rho}(\hat{U}')^\dagger]\ket{00}\\ &= \text{Tr}_\phi [\bra{00}\hat{U}'\ket{\phi}\bra{\phi} \otimes \ket{00}\braket{00} -  \braket{00}{00}\bra{00}\otimes\ket{\phi}\bra{\phi}\hat{U}'\ket{00}]\\
   &= \sum_n  [\bra{00}\bra{n}\hat{U}'\ket{\phi}\bra{\phi}\ket{n}  \ket{00}\braket{00} -  \braket{00}{00}\bra{00}\bra{n}\ket{\phi}\bra{\phi}\hat{U}'\ket{n}\ket{00}] = 0\\
\end{align*}

The only other non-trivial term in $\text{Tr}_\phi [\hat{U}' \hat{\rho}]$ is realised when $\s_j^{s_1}\s_k^{s_3}$ excites both qubits. Indeed, if $j=k$, any term with $s_1, s_3 \neq -, +$ (treated above) will annihilate $\ket{00}\bra{00}$. If $j\neq k$, $s_1 = -$ or $s_2 = -$ would annihilate $\ket{00}\bra{00}$ leaving $j \neq k$ and $s_1 = s_2 = +$ as the only non-trivial possibility. The associated term reads:
\begin{align*}
    \frac{\lambda^2}{2}M&\ket{11}\bra{00}\equiv-\frac{\lambda^2}{2}\int_{\tau_i}^{\tau_f} \mathrm{d} \tau \left(\sum_{s_2} \sum_{s_4} \left(\frac{\text{d} I_{1, + s_2}}{\text{d}\tau} I_{2, + s_4} + \frac{\text{d} I_{2, + s_2}}{\text{d}\tau} I_{1, + s_4}\right) \epsilon_{s_2 s_4}\right)  \ket{11}\bra{00}\\
    &= \frac{\lambda^2}{2}\left[ \left(I_{1, + +} I_{2, + -} + I_{2, + +} I_{1, + -}\right)-2\int_{\tau_i}^{\tau_f} \mathrm{d} \tau \left(\frac{\text{d} I_{1, + -}}{\text{d}\tau} I_{2, + +} + \frac{\text{d} I_{2, + -}}{\text{d}\tau} I_{1, + +}\right)\right]  \ket{11}\bra{00}.
\end{align*}


The next term we expand is $\text{Tr}_\phi[\hat{U}_1^2 \hat{\rho}]$. We have
\begin{align*}
    \text{Tr}_\phi[\hat{U}_1^2 \hat{\rho}] &= -\lambda^2 \sum_{j, s_1 s_2} \sum_{k, s_3 s_4} I_{j, s_1 s_2}   I_{k, s_3 s_4} \text{Tr}_{\phi}\left[\s_j^{s_1} \s_k^{s_3} \hat{a}^{s_2} \hat{a}^{s_4}  \hat{\rho} \right]\\
    &= -\lambda^2 \sum_{j, s_1 s_2} \sum_{k, s_3 s_4} I_{j, s_1 s_2}  I_{k, s_3 s_4} \langle\hat{a}^{s_2} \hat{a}^{s_4} \rangle \s_j^{s_1} \s_k^{s_3}   \hat{\rho}_i.
\end{align*}

We can simplify this result further by recalling that $\hat{\rho}_i = \ket{00}\bra{00}$. The only terms that do not annihilate the ground states of one qubit are: 
\begin{align*}
    \text{Tr}_\phi[\hat{U}_1^2 \hat{\rho}] &= -\lambda^2 \sum_{j, s_2} \sum_{k, s_4} I_{j, + s_2}  I_{k, + s_4} \langle\hat{a}^{s_2} \hat{a}^{s_4} \rangle \s_j^{+} \s_k^{+} \hat{\rho}_i-\lambda^2 \sum_{s_2} \sum_{k, s_4} I_{k, - s_2}  I_{k, + s_4} \langle\hat{a}^{s_2} \hat{a}^{s_4} \rangle \s_k^{-} \s_k^{+} \hat{\rho}_i
\end{align*}
 In parallel, we also have the conjugate terms  $\text{Tr}_\phi[\hat{\rho} (\hat{U}_1^\dagger)^2]$:
 \begin{align*}
    \text{Tr}_\phi[\hat{\rho} (\hat{U}_1^\dagger)^2] 
    &= -\lambda^2 \sum_{j, r_2} \sum_{k, r_4}  I_{j, - r_2}  I_{k, - r_4} \langle \hat{a}^{r_4} \hat{a}^{r_2}  \rangle \hat{\rho}_i \s_k^{-} \s_j^{-} 
    -\lambda^2 \sum_{r_2} \sum_{k, r_4} I_{k, + r_2}  I_{k, - r_4} \langle\hat{a}^{r_4} \hat{a}^{r_2} \rangle \hat{\rho}_i \s_k^{-} \s_k^{+}
\end{align*}

Then, we expand $\text{Tr}_\phi[\hat{U}_1 \hat{\rho} (\hat{U}_1)^\dagger]$ as follows 
\begin{align*}
    \text{Tr}_\phi[\hat{U}_1 \hat{\rho} (\hat{U}_1)^\dagger] &= +\lambda^2 \sum_{j, s_1 s_2} \sum_{k, s_3 s_4} I_{j, s_1 s_2}   I_{k, s_3 s_4} \text{Tr}_{\phi}\left[\s_j^{s_1}  \hat{a}^{s_2} \hat{\rho} \s_k^{s_3} \hat{a}^{s_4} \right]\\
    &= +\lambda^2 \sum_{j, s_2} \sum_{k, s_4} I_{j, + s_2}   I_{k, - s_4} \langle \hat{a}^{s_4}\hat{a}^{s_2} \rangle \s_j^{+} \hat{\rho} \s_k^{-}, 
\end{align*}
and finally: 
\begin{align*}
    \text{Tr}_\phi[\hat{U}_1 \hat{\rho}] &=  -i\lambda \sum_{j, s_1 s_2}  I_{j, s_1 s_2}  \text{Tr}_{\phi}\left[\s_j^{s_1}  \hat{a}^{s_2} \hat{\rho} \right]
    = -\ii\lambda \sum_{j, s_2}  I_{j, + s_2} \langle \hat{a}^{s_2} \rangle \s_j^{+} \hat{\rho}_i \\
    \text{Tr}_\phi[\hat{\rho} (\hat{U}_1)^\dagger] &= i\lambda \sum_{j, r_1 r_2}  I_{j, r_1 r_2} \langle \hat{a}^{r_2} \rangle \hat{\rho}_i \s_j^{r_1} = \ii\lambda \sum_{j, r_2}  I_{j, - r_2} \langle \hat{a}^{r_2} \rangle \hat{\rho}_i \s_j^{-}.
\end{align*}

These general results can be specialized to the context of acceleration-induced transparency by setting $I_{j, s_1 s_2} = \delta_{s_1 s_2} I_{j, s_1 s_1}$ ($I^-=0$), which leads to the update: 

\begin{align*}
    \text{Tr}_\phi[\hat{U}_1 \hat{\rho}] &= -i\lambda \sum_{j}  I_{j, + +} \langle \hat{a}^{+} \rangle \s_j^{+} \hat{\rho}_i = -i\lambda(I_{1, + +} \langle \hat{a}^{+} \rangle\ket{10}\bra{00} + I_{2, + +} \langle \hat{a}^{+} \rangle\ket{01}\bra{00})\\
    \text{Tr}_\phi[\hat{\rho} (\hat{U}_1)^\dagger] &=  i\lambda \sum_{j, -}  I_{j, - -} \langle \hat{a}^{-} \rangle \hat{\rho}_i \s_j^{-} = i\lambda (I_{1, - -} \langle \hat{a}^{-} \rangle\ket{00}\bra{10} + I_{2, - -} \langle \hat{a}^{-} \rangle\ket{00}\bra{01})\\
    \text{Tr}_\phi[\hat{U}_1 \hat{\rho} (\hat{U}_1)^\dagger] &= +\lambda^2 \sum_{j} \sum_{k} I_{j, + +}   I_{k, --} \langle \hat{a}^{-}\hat{a}^{+} \rangle \s_j^{+} \hat{\rho} \s_k^{-}\\
    &= +\lambda^2( I_{1, + +}   I_{1, --} \langle \hat{a}^{-}\hat{a}^{+} \rangle \ket{10}\bra{10} + I_{1, + +}   I_{2, --} \langle \hat{a}^{-}\hat{a}^{+} \rangle  \ket{10}\bra{01}  \\
    &\quad \quad \ + I_{2, + +}   I_{1, --} \langle \hat{a}^{-}\hat{a}^{+} \rangle \ket{01}\bra{10} + I_{2, + +}   I_{2, --} \langle \hat{a}^{-}\hat{a}^{+} \rangle \ket{01}\bra{01})\\
    \text{Tr}_\phi[\hat{U}_1^2 \hat{\rho}] &= -\lambda^2 \sum_{j} \sum_{k} I_{j, + +}  I_{k, ++} \langle\hat{a}^{+} \hat{a}^{+} \rangle \s_j^{+} \s_k^{+} \hat{\rho}_i -\lambda^2 \sum_{k} I_{k, - -}  I_{k, + +} \langle\hat{a}^{-} \hat{a}^{+} \rangle \s_k^{-} \s_k^{+} \hat{\rho}_i\\
    &= -2\lambda^2 I_{1, + +}  I_{2, ++} \langle\hat{a}^{+} \hat{a}^{+} \rangle \ket{11}\bra{00} -\lambda^2 (I_{1, - -}  I_{1, + +} + I_{2, - -}  I_{2, + +}) \langle\hat{a}^{-} \hat{a}^{+}\rangle \ket{00}\bra{00} \\
    \text{Tr}_\phi[\hat{\rho} (\hat{U}_1^\dagger)^2] &= -\lambda^2 \sum_{j} \sum_{k} I_{j,--}  I_{k,--} \langle\hat{a}^{-} \hat{a}^{-} \rangle  \hat{\rho}_i \s_k^{-} \s_j^{-} -\lambda^2  \sum_{k} I_{k, ++}  I_{k, --} \langle\hat{a}^{-} \hat{a}^{+} \rangle \hat{\rho}_i \s_k^{-} \s_k^{+}\\
    &= -2\lambda^2 I_{1,--}  I_{2,--} \langle\hat{a}^{-} \hat{a}^{-} \rangle \ket{00}\bra{11} -\lambda^2 (I_{1, - -}  I_{1, ++} + I_{2, --}  I_{2, ++}) \langle\hat{a}^{-} \hat{a}^{+}\rangle \ket{00}\bra{00}\\
    \text{Tr}_\phi [\hat{U}'\hat{\rho} +  \hat{\rho}(\hat{U}')^\dagger] &= \frac{\lambda^2}{2}M \ket{11}\bra{00} + \frac{\lambda^2}{2}M^\star \ket{00}\bra{11}.
\end{align*}
The final density matrix has the explicit expression: 
{\tiny
\begin{align*}
    \hat{\rho}_f = 
    \begin{pmatrix}
    1  - \lambda^2 (I_{1, - -}  I_{1, ++} + I_{2, --}  I_{2, ++}) \langle\hat{a}^{-} \hat{a}^{+}\rangle & +i\lambda I_{1,--} \langle \hat{a}^{-} \rangle & +i\lambda I_{2, - -} \langle \hat{a}^{-}\rangle & \frac{\lambda^2}{2}M^\star-\lambda^2 I_{1,--}  I_{2,--} \langle\hat{a}^{-} \hat{a}^{-} \rangle  \\
    -i\lambda I_{1, + +} \langle \hat{a}^{+} \rangle & \lambda^2 I_{1, + +}   I_{1, --} \langle \hat{a}^{-}\hat{a}^{+} \rangle & \lambda^2 I_{2, - -}   I_{1, ++} \langle \hat{a}^{-}\hat{a}^{+}\rangle & 0\\ 
    -i\lambda I_{2, + +} \langle \hat{a}^{+} \rangle & \lambda^2 I_{2, + +}   I_{1, --} \langle \hat{a}^{-}\hat{a}^{+} \rangle & \lambda^2 I_{2, + +}   I_{2, --} \langle \hat{a}^{-}\hat{a}^{+}\rangle & 0\\
    \frac{\lambda^2}{2}M-\lambda^2 I_{1,++}  I_{2,++} \langle\hat{a}^{+} \hat{a}^{+} \rangle & 0 & 0 & 0
    \end{pmatrix},
\end{align*}
}

The purity of $\hat{\rho}_f$ is the following:

\begin{equation}
    \textup{Tr}[\hat{\rho}_f^2]=1-2\lambda^2A_{-+}(I_{1--}I_{1++}+I_{2--}I_{2++})+o(\lambda^3),
\end{equation}

where we defined $A_{s_1s_2} = \langle \hat{a}^{s_1} \hat{a}^{s_2}\rangle - \langle \hat{a}^{s_1}\rangle \langle\hat{a}^{s_2}\rangle$. For a coherent state  $\ket{\alpha}$, $A_{-+}=1$, which means that the density matrix is indeed mixed but the purity does not depend on the coherent state, i.e. we may set it to be the vacuum state. This reproduces the previously known fact that the entanglement between the field and the state of the UDW detectors is not dependent on the coherent state, up to second order in the coupling constant \cite{Eduardo_coherent_states}. 

We have obtained the state of the two qubits $\hat{\rho}_f$ after interaction with the field, which represents our two-qubit gate. We would like to study, and optimize, the entanglement that may be acheived between the two qubits. For that purpose, we calculate the negativity as a measure of entanglement.

\subsection{Entanglement between the two qubits}\label{asec:entanglement2}

To simplify expressions, we define: $M := e^{i \theta_M} \abs{M}$, 
and $B_{s} := \langle \hat{a}^{s}\rangle$.

We start by calculating the eigenvalues of the partial transpose $\hat{\rho}_f^{\text{T}_1}$ of the density matrix $\hat{\rho}_f$ with respect to first qubit. The four eigenvalues are as follows:
\begin{align*}
    &\lambda_1=\Big[- \abs{M}+A_{-+} (I_{1--}I_{1++}+I_{2--}I_{2++})+e^{i \theta_M}A_{--} I_{1--}I_{2--}+e^{-i \theta_M} A_{++} I_{1++}I_{2++}\\
    &\!+\!\frac{|M|}{2(1+|M|)}(e^{-i \theta_M/2}B_+I_{1++}\!\!+\!e^{i \theta_M/2} B_-I_{2--})(e^{i \theta_M/2} B_- I_{1--}\!\!+e^{-i \theta_M/2} B_+ I_{2++})\Big] \frac{\lambda^2}{2} +O(\lambda^3)\\
    &\lambda_2= \Big[ \abs{M}+A_{-+} (I_{1--}I_{1++}+I_{2--}I_{2++})-e^{i \theta_M}A_{--} I_{1--}I_{2--}-e^{-i \theta_M} A_{++} I_{1++}I_{2++}\\
    &\!+\!\frac{|M|}{2(|M|-1)}(e^{-i \theta_M/2}B_+I_{1++}\!\!-\!e^{i \theta_M/2} B_-I_{2--})(e^{i \theta_M/2} B_- I_{1--}\!\!-\!e^{-i \theta_M/2} B_+ I_{2++})\Big] \frac{\lambda^2}{2} +O(\lambda^3)\\
    &\lambda_3 =\ \lambda_2\\
    &\lambda_4  =\ 0.
\end{align*}

To allow for an analysis of negativity, we need to simplify the expressions above. First we define new ladder operators $\hat{b}_1$ and $\hat{b}_2$:

\begin{align}\label{Eq:b_ladders}
    \hat{b}_1  &=\lambda I_{1,--}e^{\frac{\ii\theta_M}{2}} \left(\hat{a}-\left(1-\ii\sqrt{\frac{|M|}{|M|+1}}\right) \langle \hat{a}\rangle\right)\\
    \hat{b}_2  &=\lambda I_{2,--}e^{\frac{\ii\theta_M}{2}}\left(\hat{a}-\left(1+\ii\sqrt{\frac{|M|}{|M|+1}}\right)\langle \hat{a}\rangle\right),
\end{align}

Note that the state of the field is implicit in the definitions of these operators, hiding in the expectation value $\langle\hat{a}\rangle$. Inserting the newly defined $\hat{b}_1$ and $\hat{b}_2$ into the above expression for $\lambda_1$, it simplifies to:

\begin{equation}\label{Eq:lambda_1negativity}
     \lambda_1  = -\frac{\lambda^2}{2}|M| +\frac{1}{2}\langle (\hat{b}_1+\hat{b}_2^\dagger)(\hat{b}_1^\dagger+\hat{b}_2)\rangle +\frac{\lambda^2 I^2}{2}.
\end{equation}

For simplicity, we assumed $I_{2, s_1 s_2} = I_{1, s_1 s_2} e^{-s_2 \ i \bm{k} \cdot \bm{a}}$, i.e. a constant distance $\bm{a}$ between the two detectors, and defined $I_{1 ++} := Ie^{i\theta}$.

Crucially, we find that the second term in \cref{Eq:lambda_1negativity} is always positive. Moreover, we define a new operator $\hat{g}:= \hat{b}_1 + \hat{b}_2^\dagger$, such that the second term becomes $\frac{1}{2}\langle \hat{g}\hat g^\dagger\rangle$ which is also always positive, being the expectation value of a positive operator. Therefore the only negative term is the vacuum contribution, i.e. $-\abs{M}$:

    \begin{equation}
    \lambda_1 = -\frac{\lambda^2}{2}|M| +\frac{1}{2}\langle \hat{g}\hat g^\dagger\rangle +\frac{\lambda^2 I^2}{2}.
   \end{equation}

Similarly, we can simplify the expressions for $\lambda_2$ by defining ladder operators $\hat{b}'_1$ and $\hat{b}'_2$:
   
   \begin{align*}
       \hat{b}'_1  &= \lambda I_{1,--}e^{\frac{\ii\theta_M}{2}} \left(\hat{a}-\left(1-\ii\sqrt{\frac{|M|}{|M|-1}}\right) \langle \hat{a}\rangle\right)\\
    \hat{b}'_2  &=\lambda I_{2,--}e^{\frac{\ii\theta_M}{2}}\left(\hat{a}-\left(1+\ii\sqrt{\frac{|M|}{|M|-1}}\right)\langle \hat{a}\rangle\right).
    \end{align*}

Following the same steps as above, we find that unlike $\lambda_1$, the second eigenvalue $\lambda_2$ is always positive, and therefore does not contribute to negativity:

\begin{align*}
    \lambda_2  = &\ \frac{\lambda^2}{2}|M| +\frac{1}{2}\langle (\hat{b}^{'}_1-{\hat{b}}_2^{'\dagger})({\hat{b}}_1^{'\dagger}-\hat{b}^{'}_2 )\rangle +\frac{\lambda^2 I^2}{2},\\
    \hat{g}^{'} := &\ {\hat{b}^{'}}_1 - {\hat{b}}_2^{'\dagger},\\
    \lambda_2 = &\ \frac{\lambda^2}{2}|M| +\frac{1}{2}\langle \hat{g}^{'}{\hat{g}}^{'\dagger}\rangle +\frac{\lambda^2 I^2}{2}\geq 0.
\end{align*}

The remaining eigenvalues, $\lambda_3$ and $\lambda_4$, are non-negative as well. Thus, we found that only the first eigenvalue $\lambda_1$ can contribute to the negativity:

\begin{equation}
    \mathcal{N}=\sum_{\lambda_i<0} \abs{\lambda_i} = \abs{\lambda_1}.
\end{equation}

In order to maximize negativity, we must minimize the absolute values of the second and the third term in:

\begin{equation} \label{eq:lambda1_negativity_simple}
    \abs{\lambda_1} = \frac{\lambda^2}{2}|M| -\frac{1}{2}\langle \hat{g}\hat g^\dagger\rangle -\frac{\lambda^2 I^2}{2}.
\end{equation}

We leave the optimization of the third term to trajectory design, and focus on second term that is modulated by the state of the quantum field. We find that the amount of entanglement does not depend on the coherent state, therefore, in order to have the ability to stimulate entanglement, i.e. control the amount of entanglement created by control of the quantum field's state, we use the squeezed vacuum state of the field, given by $\hat{S}(r,\phi)|0\rangle $ where the squeezing operator is given as follows:
\begin{equation}
    \hat{S}(r,\phi) = e^{\frac{r}{2}\left(e^{-\ii \phi}\hat{a}^2-e^{\ii \phi}\hat{a}^{\dagger}^2\right)}.
\end{equation}

Here, $r$ is the squeezing parameter and $\phi$ is the squeezing angle. 
\begin{align*}
    A_{-+} &=  \langle \hat{a} \hat{a}^{\dagger}\rangle - \langle \hat{a}\rangle \langle\hat{a}^{\dagger}\rangle= 1+\sinh^2 r = \cosh^2 r\\
    A_{++} &=  \langle \hat{a}^{\dagger} \hat{a}^{\dagger}\rangle - \langle \hat{a}^{\dagger}\rangle \langle\hat{a}^{\dagger}\rangle= -\frac{1}{2}\sinh(2r) e^{-i\phi}.
\end{align*}

Therefore, for a squeezed state, 
\begin{equation}\label{Eq:gg_squeezed}
    \frac{1}{2}\langle \hat{g}\hat{g}^{\dagger}\rangle =\lambda^2 I^2\left(\cosh^2(r)-\frac{1}{2}\sinh(2r)\cos(\Theta) -\frac{1}{2}\right).
\end{equation}
By substituting the above expression in Eq.~\eqref{eq:lambda1_negativity_simple}, we get the expression for negativity for the two UDW detectors after stimulating with the squeezed vacuum state of the field:

\begin{equation}
    \mathcal{N}=\frac{\lambda^2}{2}[\abs{M}- 2I^2(\cosh^2(r)-\frac{1}{2}\sinh(2r)\cos(\Theta))].
\end{equation}

Note that the relation between $\Theta$ and $\phi$ is the following: 
\begin{equation}\label{eq:Theta}
    \Theta
 = \theta_M -2\theta + \phi + \bm{k} \cdot \bm{a}
\end{equation}
It relates the phase of the vacuum contribution $M=\abs{M} e^{i \theta_M}$, the phase $\theta$ of the time integral $I_{1,++}$, the squeezing angle $\phi$, and the phase shift $\bm{k} \cdot \bm{a}$ coming from detectors' fixed distance in trajectories.

To maximize the negativity, we need to minimize the function in \cref{Eq:gg_squeezed}. This gives us: 
\begin{equation} \label{Eq:r_optimal_squeezing}   
r_{\textup{min}} = \text{arctanh}(\cos(\Theta)).
\end{equation}

The squeezing parameter $r$ that minimizes the number operator $\langle \hat{g}\hat{g}^{\dagger}\rangle$, given the angle $\phi$ is therefore non-trivial. It corresponds to the vacuum state of the observer whose ladder operators are $\hat{g}$ and $\hat{g}^{\dagger}$. The reference frame transformation from the inertial observer to one of the detectors yields a difference in the observed number of particles for the non-inertial UDW detectors. The UDW detector sees particles in the Minkowski vacuum, with the simple expression:

\begin{equation}
    \langle \hat{g}\hat{g}^{\dagger}\rangle = \lambda^2 I^2.
\end{equation}

We may choose to not stimulate the entanglement creation by controlling the state of the field. Then we relinquish the control over how much entanglement is created, and the application of gates is not as fast as it could be. The choice of squeezed vacuum allows us some control, and, for the optimal parameters \cref{Eq:r_optimal_squeezing} reduces the expected value $\langle \hat{g}\hat{g}^{\dagger}\rangle$, and therefore enhances negativity.

\subsection{Optimization of entanglement harvesting for the calculation of a quantum computer's feasibility}

Entanglement harvesting contributes to the negativity, a term proportional to $\abs{M}$:

\begin{equation}
    M= \left(I_{1, + +} I_{2, + -} + I_{2, + +} I_{1, + -}\right)-2\int_{\tau_i}^{\tau_f} \mathrm{d} \tau \left(\frac{\text{d} I_{1, + -}}{\text{d}\tau} I_{2, + +} + \frac{\text{d} I_{2, + -}}{\text{d}\tau} I_{1, + +}\right)
\end{equation}

We will briefly compare the magnitude of $M$ to $I_+$, with the assumption that transparency condition is satisfied, therefore:
\begin{equation}
    M=-2\int_{\tau_i}^{\tau_f} \mathrm{d} \tau \left(\frac{\text{d} I_{1, + -}}{\text{d}\tau} I_{2, + +} + \frac{\text{d} I_{2, + -}}{\text{d}\tau} I_{1, + +}\right).
\end{equation}

Additionally, we will check piecewise linear trajectories, in order to more easily compare the magnitudes of $I_-$ and $I_+$. We will assume detuning, so that they are a priori of similar magnitude, and see how the time derivative contribute to the difference in M.

The $I_{j,\pm,\pm}(\tau)$ in the formula above are defined as in \cref{eq:Im_Ip_general}:

\begin{align}
    I_{j, s_1 s_2}(\tau) = \int_{\tau_i}^{\tau} \text{d}\tau' \ e^{i s_1 \Omega \tau' + i s_2(\omega t(\tau') -  \bm{k} \cdot \bm{x}_j(\tau'))}, \quad I_{j, +\pm}(\tau) \equiv I_{j}^\pm(\tau).
\end{align}

Note, that typically when we discuss $I_{\pm}$, these integrals are evaluated at boundary points, and they are not functions of proper time. The evaluated integrals are the ones to compare to, the ones dependent on time are the integrands in the expression for $M$.

Derivative acts to remove the integration: 

\begin{equation}
    \frac{d}{d\tau}I_{j, s_1 s_2}(\tau) =  \ e^{i s_1 \Omega \tau + i s_2(\omega t(\tau) -  \bm{k} \cdot \bm{x}_j(\tau))}.
\end{equation}

The integration acts to lowers some constants from the exponent: (since we are studying piecewise linear motion):

\begin{align}
    I_{j, s_1 s_2}(\tau) &= \int_{\tau_i}^{\tau} d\tau' \ e^{i s_1 \Omega \tau' + i s_2(\omega t(\tau') -  \bm{k} \cdot \bm{v}_j t(\tau'))}=\\
    &= e^{i s_1 \Omega \tau' + i s_2(\omega t(\tau') -  \bm{k} \cdot \bm{v}_j t(\tau'))}\frac{1}{i s_1 \Omega \tau' + i s_2(\omega t(\tau') -  \bm{k} \cdot \bm{v}_j t(\tau'))} \vert_{\tau_i}^{\tau}.
\end{align}

The multiplication of e.g. $I_{++}$ with $I_{-+}$ gives a cancellation in the phase function of the trajectory contribution to the total integral, and we are left with integrals of the form:

\begin{equation}
    \int d\tau' \frac{d}{d\tau'}I_{+-} I_{++} =
    \int d\tau' e^{2 i \Omega \tau'}\frac{1}{i \Omega \tau' + i (\omega t(\tau') -  \bm{k} \cdot \bm{v}_j t(\tau'))}
\end{equation}


\bibliography{references}